\documentclass[fleqn,usenatbib]{mnras}

\usepackage{newtxtext,newtxmath}

\usepackage[T1]{fontenc}
\usepackage{ae,aecompl}
\usepackage{adjustbox}
\usepackage{placeins}
\usepackage{enumitem}
\usepackage{caption}
\usepackage{comment}
\usepackage{float}
\usepackage{subcaption}
\usepackage{hyperref}
\usepackage{graphicx}
\usepackage{multirow}
\usepackage{booktabs}
\newcommand{\RNum}[1]{\uppercase\expandafter{\romannumeral #1\relax}}

\usepackage{graphicx}	
\usepackage{amsmath}	

\newcommand{\cas}{1}
\newcommand{\mpi}{2}
\newcommand{\ozgrav}{3}
\newcommand{\aut}{4}
\newcommand{\gwdata}{5}
\newcommand{\cass}{6}
\newcommand{\jbo}{7}
\newcommand{\nrao}{8}
\newcommand{\elte}{9}
\newcommand{\sarao}{10}
\newcommand{\rds}{11}
\newcommand{\ox}{12}

\title{Measurements of pulse jitter and single-pulse variability in millisecond pulsars using MeerKAT}

\author[A. Parthasarathy et al.] {A.~Parthasarathy$^{\cas,\mpi,\ozgrav}$\thanks{E-mail: aparthas@mpifr-bonn.mpg.de},
M.~Bailes$^{\cas,\ozgrav}$,
R.M.~Shannon$^{\cas,\ozgrav}$,
W.~van Straten$^{\aut}$,
S.~Os{\l}owski$^{\cas,\gwdata}$, \newauthor
S.~Johnston$^{\cass}$, 
R.~Spiewak$^{\cas,\ozgrav,\jbo}$,
D.~J.~Reardon$^{\cas,\ozgrav}$,
M.~Kramer$^{\mpi}$,
V.~Venkatraman~Krishnan$^{\mpi}$, \newauthor
T.~T.~Pennucci$^{\nrao,\elte}$,
F.~Abbate$^{\mpi}$,
S. ~Buchner$^{\sarao}$, 
F.~Camilo$^{\sarao}$,
D.~J.~Champion$^{\mpi}$,
M.~Geyer$^{\sarao}$,\newauthor
B.~Hugo$^{\sarao,\rds}$, 
A.~Jameson$^{\cas,\ozgrav}$
A.~Karastergiou$^{\ox}$, 
M.~J.~Keith$^{\jbo}$, 
M.~Serylak$^{\sarao}$
\\
$^{\cas}$Centre for Astrophysics and Supercomputing, Swinburne University of Technology, P.O. Box 218, Hawthorn, Victoria 3122, Australia \\
$^{\mpi}$ Max-Planck-Institut f\"{u}r Radioastronomie, Auf dem H\"{u}gel 69, D-53121 Bonn, Germany \\
$^{\ozgrav}$ OzGrav: Australian Research Council Centre of Excellence for Gravitational Wave Discovery.  \\
$^{\aut}$ Institute for Radio Astronomy \& Space Research, Auckland University of Technology, Private Bag 92006, Auckland 1142, New Zealand \\
$^{\gwdata}$ Gravitational Wave Data Centre, Swinburne University of Technology, P.O. Box 218, Hawthorn, VIC 3122, Australia \\
$^{\cass}$ CSIRO Astronomy and Space Science, Australia Telescope National Facility, PO~Box~76, Epping NSW~1710, Australia \\
$^{\jbo}$ Jodrell Bank Centre for Astrophysics, Department of Physics and Astronomy, University of Manchester, Manchester M13 9PL, UK\\
$^{\nrao}$ National Radio Astronomy Observatory, 520 Edgemont Road, Charlottesville, VA 22903, USA, \\
$^{\elte}$ Institute of Physics, E\"{o}tv\"{o}s Lor\'{a}nd University, P\'{a}zm\'{a}ny P. s. 1/A, 1117 Budapest, Hungary \\
$^{\sarao}$ South African Radio Astronomy Observatory, Cape Town, 7925, South Africa\\
$^{\rds}$ Department of Physics and Electronics, Rhodes University, Artillery Road, Grahamstown, South Africa, \\
$^{\ox}$ Oxford Astrophysics, Denys Wilkinson Building, Keble Road, OX1 3RH, UK
}
\date{Accepted XXX. Received YYY; in original form ZZZ}

\pubyear{2020}

\begin{document}
\label{firstpage}
\pagerange{\pageref{firstpage}--\pageref{lastpage}}
\maketitle

\begin{abstract}
Using the state-of-the-art SKA precursor, the MeerKAT radio telescope, we explore the limits to precision pulsar timing of millisecond pulsars achievable due to pulse stochasticity (jitter). We report new jitter measurements in 15 of the 29 pulsars in our sample and find that the levels of jitter can vary dramatically between them. For some, like the 2.2~ms pulsar PSR J2241--5236, we measure an implied jitter of just $\sim$ 4~ns/hr, while others like the 3.9~ms PSR J0636--3044 are limited to $\sim$ 100 ns/hr. 
While it is well known that jitter plays a central role to limiting the precision measurements of arrival times for high signal-to-noise ratio observations, its role in the measurement of dispersion measure (DM) has not been reported, particularly in broad-band observations. 
Using the exceptional sensitivity of MeerKAT, we explored this on the bright millisecond pulsar PSR J0437--4715 by exploring the DM of literally every pulse. We found that the derived single pulse DMs vary by typically 0.0085 cm$^{-3}$ pc from the mean, and that the best DM estimate is limited by the differential pulse jitter across the band. We postulate that all millisecond pulsars will have their own limit on DM precision which can only be overcome with longer integrations. Using high-time resolution filterbank data of 9 $\mu$s, we also present a statistical analysis of single pulse phenomenology. Finally, we discuss optimization strategies for the MeerKAT pulsar timing program and its role in the context of the International Pulsar Timing Array (IPTA).
\end{abstract}


\graphicspath{ {./plots/} }



\section{Introduction}
The precise monitoring of periodic radio pulses emitted from millisecond pulsars (MSPs) has enabled some of the most stringent tests of fundamental physics. It has been used to test the theory of general relativity (\citealt{Taylor_weisberg_1982, kramer_doublepulsar_2006}), alternative theories of gravity (\citealt{Zhu_alternategravity_2015,strongequiv_2020}), constrain neutron star equations of state (\citealt{ns_density2, gravitytests_4,cromartie_nanograv}), measure irregularities in terrestrial time standards (\citealt{petit_tavella_1996, Hobbs_2020_Clock}), detect planetary-mass companions (\citealt{wolszczan_frail_1992}) and can potentially be used to detect and characterise nHz-frequency gravitational radiation (\citealt{hellings_downs, foster-backer}). Precision long-term timing of an ensemble of the most stable MSPs to $<$ 100 ns root-mean-square (rms) residuals has led to placing upper limits on the stochastic gravitational wave background (\citealt{Shannon_GW_2015, lentati_ipta, GWB_Aggarwal_2019, perera_ipta_dr2_2019}), allowing constraints on the formation and evolutionary scenarios of supermassive black holes and their host galaxies (\citealt{Taylor_Simon_Sampson_2017}). 

Pulsar timing residuals, which are the differences between the observed pulse times-of-arrival (ToAs) and those predicted by a timing model, are a fundamental diagnostic tool in assessing the quality of the timing model. Numerous studies since the 1970s have shown that the scatter in timing residuals are larger than that expected from the formal uncertainties (i.e., the uncertainties reported from a match-filtered based arrival time determination algorithm) alone (\citealt{groth_crab_1975,cordes_downs_1985,oslowski_2011,Shannon_jitter_2014,lam_jitter_2019}). This excess noise can be categorised into a time-correlated, red-noise component and an uncorrelated white-noise component. One of the main contributing sources to the red noise is caused by rotational irregularities in the pulsar's spin period also known as spin noise or timing noise (\citealt{boynton_timingnoise, cordes_noise_1980}). Spin noise manifests as a low-frequency process in pulsar timing residuals over timescales of months to years. Many studies have attempted to characterise the strength and non-stationarity of spin noise across the pulsar population (\citealt{shannon_cordes_2010,psj1}) finding that although it is widespread in pulsars, it is weaker in millisecond pulsars (\citealt{lam_msp}). Additional contributions to the red-noise component can arise from quasi-periodic processes, due to magnetospheric torque variations (\citealt{kramer_switching_2006, lyne_switching_2010}), unmodelled planetary companions (\citealt{Shannon_asteroid_2013, Kerr_planetes_2015}), unmodelled dispersion measure (DM) variations (\citealt{Keith_ism_2013}), turbulence in the interstellar medium (\citealt{cordes_shannon_2010, lam_msp}) and uncertainties in the solar system ephemeris (\citealt{ssb_masses,Caballero_ipta_2018}). 

On timescales of minutes to hours, the excess noise in the timing residuals is typically dominated by the uncorrelated white-noise component. This excess white noise, in addition to radiometer noise, can arise from several sources, the most significant of which is from differences between the integrated pulse profile (the averaged phase-resolved light curve of the pulsar) and a template profile (average of a finite number of pulses). This difference contributes directly to the excess white noise which results in observed ToA uncertainties being higher than the predicted formal values.  The formal uncertainty in the arrival time is derived from the template-matching algorithm, which models the integrated pulse profile ($P$) as a scaled (by a factor $A$) and offset (by a constant $B$) version of the template ($O$) rotated by a phase shift $\phi$ with additional white noise $N(t)$, expressed as (\citealt{Taylor_timing}),
\begin{equation}
P(t)=A O(t-\phi)+B+N(t).
\end{equation}

Observations of single pulses from pulsars with high flux densities have exhibited variations in their pulse morphology along with correlated variations in their phase and amplitudes (\citealt{drake_singlepulse_1968, psr_integratedprofile, singlepulsenoise_4}). Aside from the fact that pulse profile variability adversely affects the attainable precision in pulsar timing experiments with MSPs, it is important to acknowledge that the general phenomenology has been extensively studied across the pulsar population on both short (seconds to hours) and long (months to years) timescales. It has been known that emission changes can occur in pulsars on short timescales (\citealt{modechanging_backer,nulling}); where pulse profiles were observed to switch between two or more distinct emission states (known as \textit{mode changing}) or where the profile was either in a weak emission state or completely turned `off' (known as \textit{nulling}). Pulsars categorized as \textit{intermittent} have been observed to cycle between quasi-periodic phases in which the radio emission is either clearly present or invisible (\citealt{kramer_switching_2006, camilo_nulling_2012, lyne_intermittent_2017}) and in all cases the observed emission changes have been correlated with the pulsar's rotational behaviour. Changes in pulse morphology have also been attributed to precession of the pulsar's spin axis, which causes different regions of the emission beam to orient along our line-of-sight (\citealt{weisberg_precession_1989,Kramer_1998_precession}). In an important study, \cite{cordes_downs_1985} analysed 24 pulsars and concluded that pulse shape variations or jitter were significant in a large fraction of their sample and proposed that it is likely to occur in all pulsars with varying degrees of importance. 

Profile changes in MSPs have been reported in very few cases. \cite{hotan_J1022_2004} \& \cite{willem_polcal2} showed that previous reports of profile instabilities in PSR J1022+1001 (\citealt{kramer_1999_msp}) are possibly due to instrumental polarimetric calibration errors while a recent study suggests that calibration alone cannot account for the observed profile variations (\citealt{prajwal_1022}). \cite{Shannon_1643_2016} reported a broad-band profile change in PSR J1643--1224 (also observed by \citealt{brook_profilechanges}) which may have been due to a disturbance or a state change in the pulsar magnetosphere. Very recently, the brightest and closest MSP, PSR J0437--4715 exhibited a significant change in its integrated pulse profile (\citealt{Kerr_2020_IPTA}), indicating that such abrupt profile changes may be common among MSPs as well. In contrast to observations of young pulsars, the profile changes in both MSPs were not accompanied with measurable changes in their spin down rate. 

A few studies have recently focused on studying jitter noise in MSPs and its effect on limiting the attainable precision through pulsar timing (\citealt{oslowski_2011,liu_jitter_2012,Shannon_jitter_2014,lam_jitter_2019}). Jitter noise is a stochastic process common to all pulsars, arising from intrinsic self-noise in the pulsar emission mechanism and is thought to be a wide-band phenomenon (\citealt{taylor_intensity_1975,singlepulsenoise_2}). Since single pulse morphology changes stochastically from pulse to pulse, and can be measurable in high signal-to-noise (S/N) observations, the averaged pulse profile ($P$) will have a shape that is different from the template, thus causing an excess scatter in the measured ToA uncertainty. This excess scatter can be measured from its contribution to the rms of the ToAs, expressed as $\sigma_{\rm J}(N_{\rm p})$, where $N_{\rm p}$ is the number of averaged pulses or can also characterised with a dimensionless jitter parameter ($f_{\rm j}$), as the ratio of $\sigma_{\rm J}(N_{\rm p})$ to the pulse period ($P$). Unlike the formal ToA uncertainty ($\sigma_{\mathrm{S} / \mathrm{N}}$), $\sigma_{\rm J}$ is independent of the S/N. Furthermore, additional scatter in the ToAs can also result from narrow-band diffractive interstellar scintillation (DISS) along with the frequency dependence of the pulse profile, especially if the profiles are averaged over large bandwidths \cite[][]{nanograv,Shannon_Cordes_2017_DISS}.  A second effect related to diffractive scintillation, is caused by stochasticity in the pulse broadening and has been termed the finite-scintle effect \cite[][]{cordes_J1937_1990,cordes_shannon_2010}. This effect can be is greater for pulsars which show larger degrees of scatter broadening, so is especially important for high dispersion-measure pulse observed at lower frequencies. 


\cite{oslowski_2011} analysed 25 hours of high-precision timing data of PSR J0437--4715 (using the Murriyang/64-m Parkes radio telescope, at an observing frequency of $\sim$ 1400 MHz) reporting that in one hour ($N_{\rm p}$ $\sim$ $10^{6}$ pulses), pulse jitter limits the attainable timing precision to $\sim$ 30 ns. \cite{shannon_cordes_J1713_2012} similarly reported the jitter in PSR J1713+0747 to be $\sim$ 20 ns in an hour. Jitter measurements for a sample of 22 MSPs as part of the Parkes Pulsar Timing Array (PPTA; \citealt{ppta}) project were reported by \cite{Shannon_jitter_2014} with PSR J1909--3744 showing the lowest levels of jitter noise of $\sim$ 10 ns in an hour. They modelled the contribution of jitter as a function of observing time ($T$) as $\approx$ $0.5 W_{\rm eff} \sqrt{P/T}$. More recently, \cite{lam_jitter_2019} detected jitter in 43 MSPs as part of  timing program of the North American Nanohertz Observatory for Gravitational Waves \cite[NANOGrav, ][]{nanograv_11yr} and found significant frequency dependence of jitter in 30 of them. These studies clearly show that pulse jitter is a generic property of MSPs, that it dominates the white noise budget when observing with increased S/N and that robust characterisation of jitter noise is vital in the search for stochastic nHz gravitational wave background using pulsar timing arrays. 

The MeerKAT radio telescope is several times more sensitive than the Parkes radio telescope and offers the opportunity to detect and constrain jitter in many more MSPs at declinations $<$ $+$40$^{\circ}$ . In Section \ref{observations_sec} we describe the MeerTime MSP observing program, the relevant data processing by the observing backend and the various processing steps implemented in the data reduction pipeline. We discuss the methodology used to estimate jitter, report jitter measurements for 29 MSPs and discuss its frequency dependence in Section \ref{jitter_measurements_sec}. In Section \ref{0437_DM} we discuss fundamental limits imposed on DM measurements in PSR J0437--4715. Statistical studies of single pulses allow us to link timing variations and shape variations thus providing further insights into characterising jitter noise. In Section \ref{singlepulse_sec}, we describe the statistical properties of MSP single pulses and discuss these results in the context of timing variations. Finally, in Section \ref{conclusion_sec}, we discuss the implications of our results and the role of the MeerTime Pulsar Timing Array (MPTA) programme in aiding high precision pulsar timing and PTA experiments.

\section{Observations and data reduction} \label{observations_sec}

The MeerKAT radio telescope is the South African precursor for the Square Kilometre Array (SKA) mid radio telescope located in the Great Karoo region, and is capable of observing the large population of known pulsars in the Southern and Northern hemispheres. The MeerTime collaboration (\citealt{Bailes_2020}) is one of the MeerKAT Large Survey Projects, which has been provisionally awarded many months of observing time. The MPTA is one of the four major science themes as part of MeerTime, which focuses on the precision timing of MSPs and is poised to contribute to an internationally coordinated effort to detect the stochastic gravitational wave background using pulsar timing arrays through the International Pulsar Timing Array (IPTA, \citealt{hobbs_ipta_2010,verbiest_ipta_2016,perera_ipta_dr2_2019}). The current observing strategy for the MPTA is to attain sub-microsecond formal timing precision on as many pulsars as possible with integration times less than 2048 seconds. For each pulsar, the integration time is set to match the median expected time either based on previous observations from the MeerKAT MSP census program or previous monitoring campaigns. A maximum integration time of 2048 seconds is imposed per epoch and a minimum integration time of 256 seconds is used if sub-microsecond timing precision is achievable in less than that time. The MPTA, since commencing observations from February 2019, attains timing precision of $<$ 1$\mu s$ on $\sim$ 70 MSPs in a total integration time of approximately 11 hours. The Parkes Pulsar Timing Array, in comparison, achieves sub-microsecond precision in 22 MSPs in 24 hours and NANOGrav achieves the same   precision on 47 MSPs (\citealt{NG_2020a,NG_2020b}).

For the jitter analysis presented here, we selected MSPs observed with MeerKAT that exceeded a S/N per pulse of unity following \cite{shannon_cordes_J1713_2012}. These included $\sim$ 350 observations of 29 pulsars over a span of about a year (starting from February 2019). Approximately 80\% of these observations had an integration time of $<$ 900 seconds. We used the L-band receiver with a system temperature of $\sim$ 18~K, operating at a frequency range between 856 MHz and 1712 MHz. The dual-polarization and channelized signal from the beamformer were processed in the Pulsar Timing User Supplied Equipment (PTUSE) using the \textsc{dspsr}\footnote{\url{http://dspsr.sourceforge.net/}} software library (\citealt{atnf_pks_swin_rec}). \textsc{dspsr} provides two signal processing pipelines and produces fold-mode and search-mode data products. Pulsar timing applications use the fold-mode pipeline to produce frequency- and phase-resolved averages of the polarised flux for the target pulsar. Search-mode applications (for single pulses) on the other hand, use \textsc{digifits} to produce high time resolution spectra (filterbanks) data products. A detailed description of the MeerKAT pulsar timing infrastructure including polarisation calibration is provided in \citealt{Bailes_2020}. For the analysis presented here however, we only used the total intensity profiles.

The coherently dedispersed folded archives and search-mode data products produced by PTUSE are automatically ingested by the MeerKAT \textit{kat-archive} and transferred to the OzStar HPC facility at Swinburne University of Technology through an authenticated download for post-processing. PTUSE produces folded archives of 8 second integrations with 4 Stokes parameters, 1024 frequency channels across the observing bandwidth with 1024 phase bins. Although the receiver has a bandwidth of 856 MHz, a portion of the bandwidth is ignored due to bandpass filter roll-off which reduces the usable bandwidth to 775.75 MHz and thus the number of frequency channels to 928. These archives are further processed by the fold-mode processing pipeline \textsc{meerpipe}\footnote{https://bitbucket.org/meertime/meerpipe/src/master/}, which generates RFI-excised full-frequency and time resolution \textsc{psrfits} based archives (\citealt{psrchive_psrfits}) along with decimated products with user-specified frequency and time resolution and associated ToAs.

The fragmented 8~second folded archives are integrated using \textsc{psrchive} tool  \textit{psradd} and aligned using an up-to-date pulsar ephemeris. The ephemerides are automatically checked using a series of pre-defined standards which is then followed by a manual vetting process. These are automatically version controlled and used by PTUSE , ensuring accountability.  The noise-free templates used for timing are generated either manually (using  the \textsc{psrchive} tool  \textit{paas}) or automatically wavelet-smoothed using the  \textsc{psrchive} tool  \textit{psrsmooth}. Frequency-dependent template profiles (or portraits) are generated for pulsars which show significant profile evolution across frequency using \textsc{PulsePortraiture}\footnote{https://github.com/pennucci/PulsePortraiture} (\citealt{tim_wideband}).

To study single pulses, the filterbanks are recorded at a time-resolution of 9$\mu s$ across 768 frequency channels and are stored in \textsc{psrfits} format. For the analysis presented here, the post-processing of single pulses is done using \textsc{dspsr}.


RFI excision is implemented using a modified version of \textsc{coastguard} (\citealt{Lazarus_coastguard_2016}) which uses a template profile to identify the phase-bins containing the pulsar signal and computes profile residuals by subtracting the observed profile from the template. The folded data cube with full-frequency and time resolution after integrating the four  Stokes parameters is used to excise RFI. Using  metrics as described in \cite{Lazarus_coastguard_2016}, RFI mitigation is performed on the Fourier transform of the profile residuals. For the analysis presented here, only frequency integrated template profiles are used for RFI excision. Furthermore, manual inspection of the output archives is performed to ensure that no residual RFI is present in the data. 

The reference signal produced by the incoherent sum of the noise diode signals in the MeerKAT array cannot be used to calibrate the absolute gain, differential gain and differential phase of the system as it deviates from 100\% linear polarization (\citealt{Bailes_2020}). However, the differential gain is calibrated to within 1\% during the procedure that is used to phase up the tied array, leaving only the absolute gain and differential phase to be calibrated.  The absolute gain can be calibrated using a separate set of flux calibration observations of a radio source known to have constant flux density, such as PKS~B1934$-$638 and in a subset of MeerTime observations, the differential phase also happens to be very close to zero.  Therefore, for these observations, it is sufficient to perform only the feed hand (basis) and parallactic angle (projection) corrections which are implemented in the processing pipeline using \textit{pac}.

The cleaned and vetted archives are then decimated into various data products containing different number of frequency channels and sub-integrations. The ToAs are then calculated by cross correlating these sub-banded observations with a frequency integrated template profile in the Fourier domain. The formal uncertainties produced by this method assumes that the only source of noise in the measurement is white radiometer noise and thus, underestimates the true ToA uncertainty. The meta-data associated with each ToA follows the IPTA convention (\citealt{verbiest_ipta_2016}). Wideband ToAs that account for frequency-dependent profile evolution are also produced (\citealt{widebandtiming_1}). For the results presented here, we used profiles that are frequency-averaged to 32 channels with 8~second subintegrations. 

\section{Jitter measurements} \label{jitter_measurements_sec}

In this section we first describe the methodology used to estimate jitter using frequency averaged ToAs followed by the use of wideband templates to examine the frequency dependence of jitter. 

\subsection{Methodology}

The sub-banded ToAs produced by \textsc{meerpipe} are fitted to curated pulsar ephemerides to obtain sub-banded timing residuals. Since the observations typically have a 5 minute duration, only the spin-frequency ($\nu$) and DM are fitted for and only ToAs derived from 8~s profiles with S/N $>$ 10 are retained in the analysis\footnote{We have cross-checked these measurements with a wideband template as well and did not find any discrepancies.}. To compute the rms uncertainty of jitter in $T_{\rm {sub}}$($=8$s), we frequency average the sub-banded timing residuals and compute the quadrature difference of the rms of the frequency-averaged residuals and the rms expected from ideal simulated data sets as expressed in equation,

\begin{equation} \label{jitter_eq}
\sigma_{\rm J}^{2}\left(T_{\rm{sub}}\right)=\sigma_{\mathrm{obs}}^{2}\left(T_{\rm {sub}}\right)-\sigma_{\mathrm{sim}}^{2}\left(T_{\rm {sub}}\right).
\end{equation}

The uncertainties from the observed, frequency-averaged ToAs are induced in an idealised ToA data set to generate the simulated data. Using the  {\sc tempo2} {\sc fake} tool (\citealt{hobbs_tempo2_2006}),  we simulate  $\sim$ 1000 realisations of the data set for each pulsar (for every epoch) to obtain the variance of $\sigma_{\mathrm{sim}}^{2}\left(T_{\rm {sub}}\right)$. We assume that all excess noise observed in the arrival time measurements are caused due to jitter since other effects that lead to such short-timescale perturbations vary more strongly with observing frequency and cause perturbations in the ToAs on longer ($\sim$ hours) timescales than what is reported here (\citealt{shannon_cordes_J1713_2012}). Distortions in the pulse profile caused due to imperfect polarization calibration typically tend to vary with the parallactic angle of the receiver and are caused on much longer timescales than a few minutes. The methodology presented here is similar to that discussed in \cite{Shannon_jitter_2014}, except that we use simulated ToAs rather than simulated profiles. Since jitter is expected to scale proportionally to $1/\sqrt{N_{\rm p}}$, where $N_{\rm p}$ is the number of pulses, we can estimate the implied jitter in one hour to be,

\begin{equation} \label{implied_jitter_eq}
\sigma_{\rm J}\left(\rm {1 hour}\right) = \sigma_{\rm J}\left(T_{\rm{sub}}\right)/\sqrt{3600/\left(T_{\rm{sub}}\right)}.
\end{equation}

We also use the Bayesian pulsar timing package, \textsc{temponest} (\citealt{temponest}) as a consistency check for our jitter measurements. To estimate jitter using \textsc{temponest}, we determine the standard white noise parameter, EQUAD, used commonly in pulsar timing analyses\footnote{Since we use frequency-averaged ToAs, EQUAD and ECORR result in similar estimates of jitter noise.  The ECORR parameter models short-timescale noise processes that result in correlated sub-banded TOAs within an epoch/observation, but which are otherwise uncorrelated between epochs/observations. \cite[][]{nanograv_9y}. }. EQUAD represents a source of time-independent noise which could arise from stochastic shape variations in the integrated pulse profile. 

We claim to have detected jitter in a pulsar if $\sigma_{\mathrm{obs}}^{2}\left(T_{\rm {sub}}\right)$ is greater than 95\% of the simulated $\sigma_{\mathrm{sim}}^{2}\left(T_{\rm {sub}}\right)$ values, which suggests that the rms of the frequency-averaged residuals is higher due to excess scatter in the observed ToAs than that caused due to radiometer noise.

\subsection{Jitter measurements from frequency-averaged ToAs}

Table \ref{jitter_table} reports the jitter measurements for 29 MSPs in our sample using frequency averaged ToAs. We constrain jitter in 13 pulsars and report upper limits for the remaining. Pulsars with either new jitter measurements or new upper limits are highlighted in the table. PSR J2241--5236 has the lowest level of jitter hitherto reported. Our measurements are consistent within uncertainties to previously published values which are also reported in the table.

\begin{figure*}
\centering
\includegraphics[angle=0,width=0.75\textwidth]{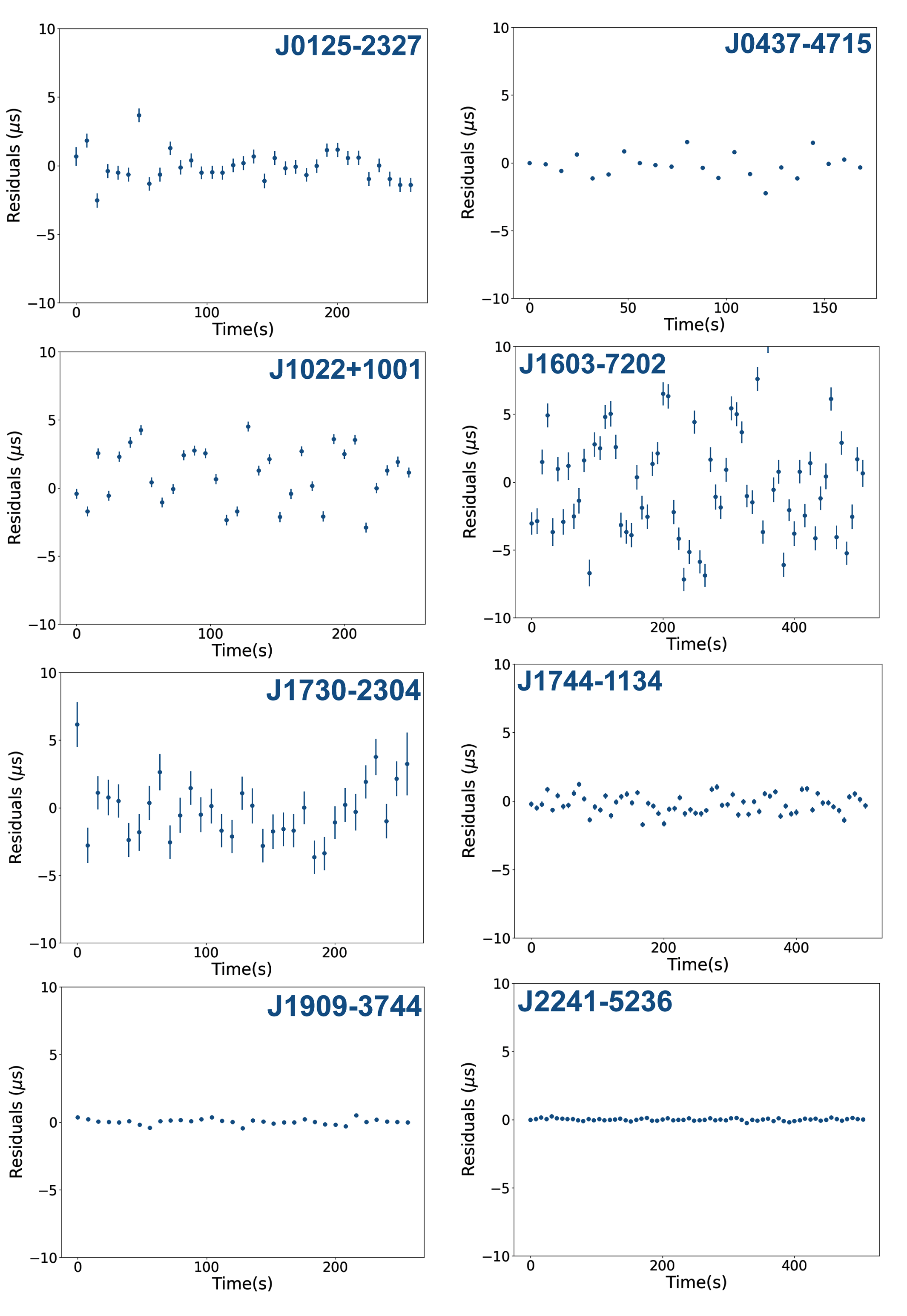}
\caption{\label{jitter_avg_plots}
Frequency averaged timing residuals for eight pulsars from our sample showing varying levels of jitter. All y-axes are plotted on the same scale for ease of comparison. }
\end{figure*}

In Figure \ref{jitter_avg_plots}, we show a representative sample of frequency averaged timing residuals of eight pulsars with different levels of jitter. To highlight the markedly different strengths of jitter noise across the population, the timing residuals are all plotted on the same y-scale. 

\begin{table*}
\caption{\label{jitter_table} Jitter measurements and upper limits for 29 MSPs in our sample. For each pulsar, the parameters reported here are corresponding to the brightest observation, i.e, with the highest average S/N per pulse. For reference, the median S/N per pulse computed from all selected observations per pulsar is also reported. Columns two and three report the period and the DM while columns six to eight report the integration time, the mean ToA error and the weighted RMS values in 8~s sub-integrations. The last two columns report the implied jitter in one hour and a reference to a previously published jitter measurement where available. S+2014 refers to \protect\cite{Shannon_jitter_2014} and L+2019 refers to \protect\cite{lam_jitter_2019} 
(values are scaled from single-pulse rms values). Pulsar names in bold represent either a new detection or a new upper limit.}

\centering
\vspace{2ex}     
\begin{tabular}{lrrrrrrlll}
\hline
\hline
 PSR &  P & DM  &S/N$_{\rm pulse, max}$ &  S/N$_{\rm pulse,med}$ & $\rm{T}_{\rm obs}$ & $\sigma_{\rm ToA}$ & WRMS & Implied $\sigma_{\rm J}(\rm {hr})$ & Previous $\sigma_{\rm J}(\rm {hr})$ \\
  & (ms)  &  (pc\,cm$^{-3}$) & & & (s) & ($\mu$s) & ($\mu$s) & (ns) & (ns) \\
\hline
J0030+0451 & 4.9 & 4.4           & 1.9   &  1.4   & 370   & 0.298  & 1.765  & $<$60 & 60$\pm$5 (L+2019) \\
\textbf{J0125--2327}  & 3.7 & 9.6   & 3.6  & 3.6   & 256  & 0.183  & 1.108 & 48$\pm$13 & -  \\
J0437--4715 & 5.8 & 2.6                 &  139  & 112  & 180 & 0.019 & 0.868 & 50$\pm$10 & 48.0$\pm$0.6 (S+2014)   \\
\textbf{J0636--3044}  & 3.9 & 16.0            & 1.6  & 1.1  & 1170 & 1.285 & 3.555 & 100$\pm$30 & -  \\
J0711--6830 & 5.5 & 18.4                     &  2.1  & 1.6  & 256  & 0.619 & 1.284  & 60$\pm$20 & $< 90$ (S+2014)  \\
\textbf{J0900--3144} & 11.1 & 75.7  &   2.9  &  2.5  & 256  & 0.554 & 3.687 &  $<$130 & -  \\
J1017--7156  & 2.3 & 94.2                    &   2.4  & 2.2  & 256  & 0.149 & 0.259 & $<$10 & $<100$ (S+2014)  \\
J1022+1001  & 16.5 & 10.3                    &   23.3 & 5.6  & 256 & 0.167 & 2.054 & 120$\pm$20 & 280$\pm$140 (S+2014), 265$\pm$20 (L+2019)  \\
J1024--0719 & 5.2 & 6.5                     & 2.1   &  1.8  & 256  & 0.204  & 0.886 & $<$30 & 18$\pm$10 (L+2019) \\
\textbf{J1045--4509}  & 7.5 & 58.1          &  2.9  & 2.0  & 256  & 0.555 & 3.368 & 130$\pm$75 & $< 900$ (S+2014)  \\
\textbf{J1157--5112}  & 43.6 & 39.7           &   1.7  & 1.5  & 3064  & 3.675 & 32.848 & $<$690 & -   \\
J1600--3053  &3.6  & 52.3                     &   1.9  & 1.4 & 256  & 0.118 & 0.711 & $<$30 & $<200$ (S+2014)  \\
J1603--7202 & 14.8 &38.0                     &   8.1  & 3.8   & 512 & 0.297 & 3.947 & 180$\pm$40 & 300$\pm$56 (S+2014)  \\
\textbf{J1622--6617} & 23.6 &87.9            &   1.6  & 1.5  & 256  & 2.158 & 26.291 & $<$300 & -   \\
\textbf{J1629--6902} & 6.0 & 29.5             & 1.5  & 1.4  & 256  & 0.331 & 1.917 & $<$60 & -  \\
J1643--1224 & 4.6 & 62.4                    &   2.4  & 2.0 & 256  & 0.303 & 1.647 & $<$60 & $<500$ (S+2014), 31$\pm$12 (L+2019)  \\
\textbf{J1730--2304} &  8.1 & 9.6           &  3.0 & 1.7 & 256 & 0.451 & 1.973 & 80$\pm$45 & $<400$ (S+2014)  \\
J1744--1134     & 4.1 & 3.1               &  6.6  & 2.6  & 512  & 0.026 & 0.686 & 30$\pm$6 & 37.8$\pm$0.8 (+2014), 44$\pm$1 (L+2019)  \\
\textbf{J1756--2251} & 28.5 & 121.2             &   2.1  & 1.7  & 1345  & 0.926 & 21.176 & $<$500 & -   \\
\textbf{J1757--5322} & 8.9 & 30.8            &   1.9  & 1.4  & 1242  & 0.704 & 4.196 & 130$\pm$45 & -  \\
\textbf{J1802--2124} & 12.6 &149.6            &   1.3  & 1.4  & 256  & 0.545 & 8.485 & $<$80 & -  \\
J1909--3744  & 2.9 & 10.4                    &   9.6 & 3.2  & 256  & 0.021 & 0.199 &  9$\pm$3 & 8.6$\pm$0.8 (S+2014), 14$\pm$0.5 (L+2019)  \\
\textbf{J1918--0642} & 7.6 & 26.5            &   2.1  & 2.1  & 512  & 0.681 & 1.645 &  $<$55 & -   \\
\textbf{J1946--5403} & 2.7 &23.7           & 1.2 & 1.2 & 256  & 0.090  & 0.359 & $<$9 & -  \\
J2010--1323 & 5.2 & 22.2                     &  1.5 & 1.5  & 256  & 0.224 & 1.048 &  $<$80 & 59$\pm$3 (L+2019)   \\
\textbf{J2039--3616} & 3.3 & 24.0            &   1.1 & 1.1  & 512 & 0.166 & 0.937 &  $<$25 & -   \\
J2129--5721& 3.7 & 31.9 & 3.3  & 3.3  & 360  & 0.285 & 0.493 &  $<$11 & $< 400$ (S+2014)  \\
J2145--0750 & 16.1 & 9.0 & 11.1 & 3.6  & 256 & 0.331 & 3.883 &  200$\pm$20 & 192$\pm$6 (S+2014), 173$\pm$4 (L+2019) \\
\textbf{J2241--5236} & 2.2 & 11.4           &  10.3 & 2.3  & 512  & 0.011 & 0.086 & 3.8$\pm$0.8 & $< 50$ (S+2014) \\
\hline
\end{tabular}
\end{table*}

\subsubsection{Contribution of scattering noise to the jitter measurements}

The other short term noise source that can contribute to excess white noise is related to propagation in the interstellar medium. Stochasticity in the pulse-broadening function, referred to as the finite-scintle effect \cite[][]{cordes_J1937_1990,cordes_shannon_2010,lam_16_intraday} will cause arrival time variations that can be significant for highly scattered pulsars.  The strength of this effect depends on the pulse broadening time $\tau$, and the number of scintles in the observation $N_s$, $\sigma_{\rm FS} = \tau/\sqrt{N_s}$. 
The numbers  of scintles is $N_s = (1 + \eta \Delta \nu/\nu_d)(1+ \eta \Delta T/t_d)$,  
where $\Delta T$ and $\Delta \nu$ are the observing time and bandwidth while $t_d$ and $\nu_d$ are the diffractive scintillation time and bandwidth, 
and $\eta \approx 0.3$ is the scintillation filling factor \cite[][]{cordes_shannon_2010}.
Of the pulsars for which we have detected excess white noise, only one, PSR~J1045$-$4509 has a dispersion measure above $50$\,pc\,cm$^{-3}$. Based upon its diffractive scintillation bandwidth and time scale   we expect that the contribution of the finite-scintle effect to be $40$~ns, which is lower than the measured level of jitter noise.  
The high dispersion measure pulsar PSR J1017$-$7156 has a tight constraint on excess noise $\sigma_{\rm J} < 10$\,ns.  This pulsar however has an under-turbulent line of sight with a scintillation bandwidth of $\Delta \nu \approx 2$\,MHz and time scale of $10$\,min at 1.4\,GHz (\citealt{Coles_ESE}). As a result, the strength of the effect is estimated to be $\sim$ 5 ns in a 1\,h observations, consistent with our jitter limit. We defer further analysis of this effect to future work, noting that studies of the effect would be particularly amenable with the MeerKAT UHF system.

\subsection{Frequency dependence of jitter}

By utilizing the large fractional bandwidth of MeerKAT, we are able to study the frequency dependence of jitter in our sample of MSPs. Based on the S/N of the observation we generate ToAs per sub-band to compute $\sigma_{\rm J}$ as a function of frequency channel. We do not detect significant frequency dependence of jitter in any other pulsars except in PSR J0437--4715. Higher S/N observations of these pulsars, especially during scintillation maxima might enable such studies. 

For PSR J0437--4715, we find that jitter decreases with increasing observing frequency. In the lower part of the band at $\sim$ 910 MHz, we measure $\sigma_{\rm J}$ to be 63$\pm$25~ns, at $\sim$ 1300 MHz, we measure $\sigma_{\rm J}$ to be 50$\pm$15~ns, while at $\sim$ 1660 MHz, we measure $\sigma_{\rm J}$ to be only 24$\pm$20~ns. This is consistent with \cite{Shannon_jitter_2014}, who reported jitter to be modestly greater at lower frequencies. However, it is important to note that PSR J0437--4715 shows significant frequency-dependent profile evolution which could likely bias the uncertainties on the jitter measurements at lower and higher bands due to template fitting errors. To account for these limitations, we generate a wideband template that models the frequency evolution of the profile.  

\begin{figure}
\centering
\includegraphics[angle=0,width=1\columnwidth]{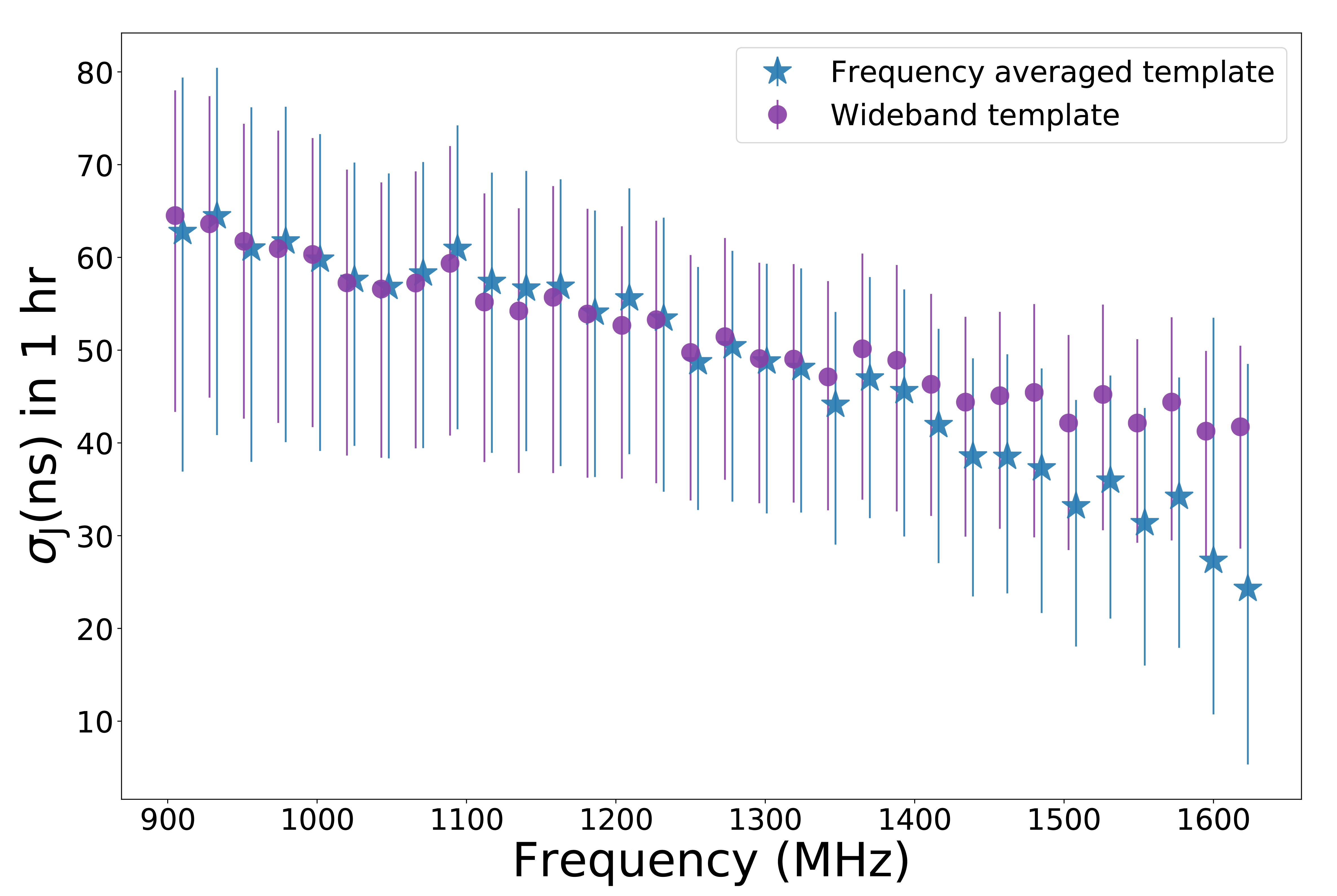}
\caption{\label{0437_jitter_frequencydependence}
Jitter as a function of observing frequency for PSR J0437--4715, using ToAs generated from frequency-averaged (blue) and wideband templates (purple). The frequency-averaged points are offset by 5 MHz for clarity. The observing bandwidth is averaged to 32 frequency channels. There is a moderate dependence of jitter on observing frequency. }
\end{figure}

Using \textsc{PulsePortraiture} and the methodology described in \cite{tim_wideband}, we create a frequency-dependent smoothed template for PSR J0437--4715. Using this, we compute sub-banded ToAs with uncertainties that account for the frequency-dependent profile evolution. Figure \ref{0437_jitter_frequencydependence} shows the frequency dependence of jitter using frequency-averaged and wideband templates. Using these ToAs, at $\sim$ 910 MHz, we measure $\sigma_{\rm J}$ to be 64$\pm$20~ns, at $\sim$ 1300 MHz, we measure $\sigma_{\rm J}$ to be 50$\pm$13~ns, while at $\sim$ 1660 MHz, we measure $\sigma_{\rm J}$ to be 42$\pm$12~ns. The deviations in the measurements of $\sigma_{\rm J}$ using frequency-averaged and sub-banded ToAs are consistent within uncertainties. It must be noted that as the template deviates from an accurate description of the profile, we get an artificial lower value of jitter. The varying levels of jitter in higher and lower frequency bands in PSR J0437--4715 can most likely be attributed to the narrowing of the pulse profile at higher frequencies.

\begin{figure} 
\centering
\includegraphics[angle=0,width=1\columnwidth]{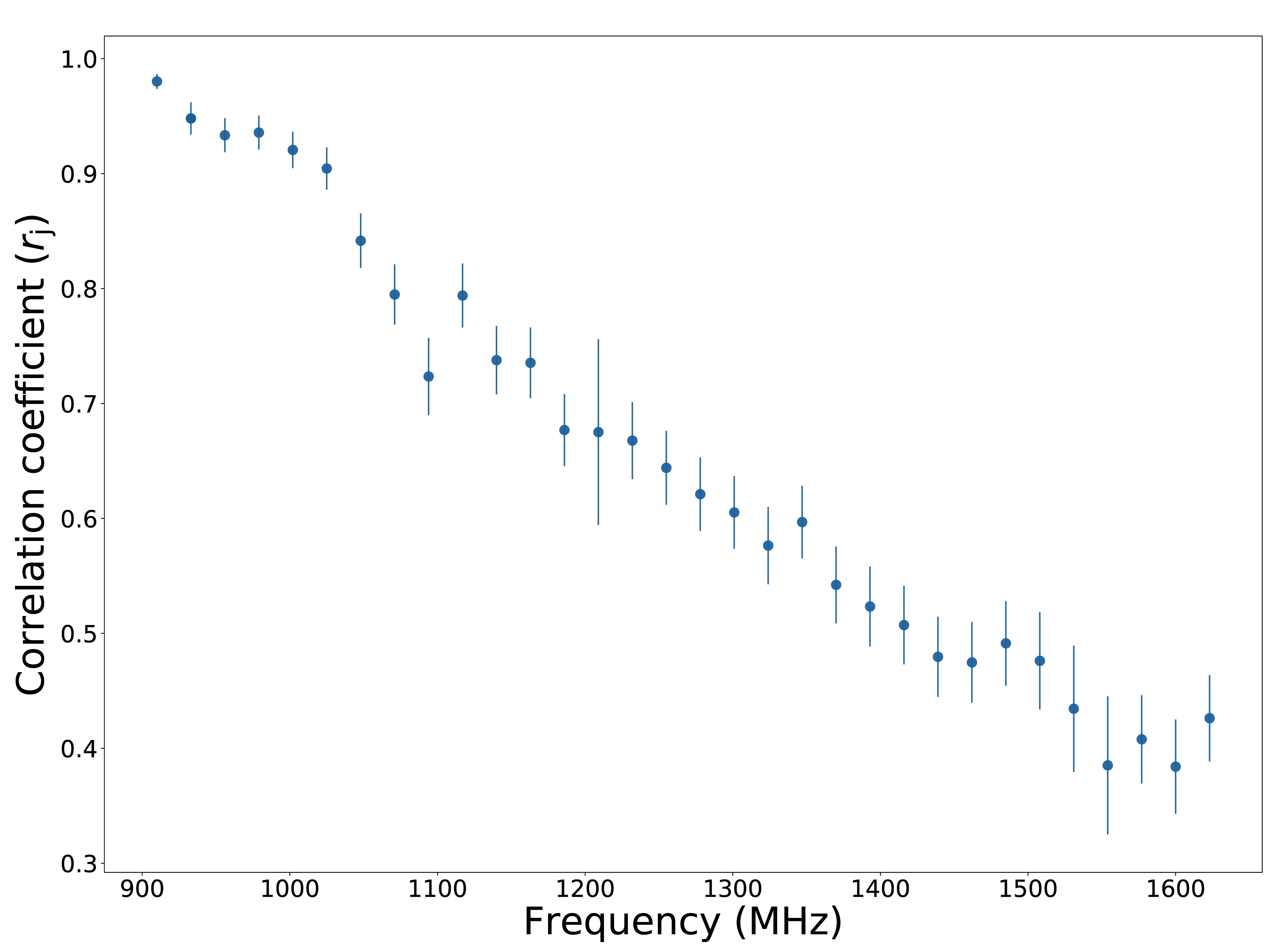}
\caption{\label{jitter_channelsep}
ToA correlation as a function of observing frequency (in MHz) for PSR J0437--4715 showing the increasing decorrelation of jitter with increasing channel separation. The lowest frequency channel at a frequency of $\sim$ 900 MHz is considered as the reference frequency.}
\end{figure}

Owing to our ability to measure jitter in individual sub-bands and also due to the high flux density of the pulsar, we can estimate the degree of correlation of jitter between the bands. We see a high degree of correlation ($\sim$0.9) between adjacent frequency bands, while the degree of correlation significantly reduces ($\sim$ 0.4) between the lowest and highest frequency bands. Assuming a reference frequency of $\sim 910$ MHz, we compute the correlation strength ($r_{\rm j}$) as a function of increasing channel separation as shown in Figure \ref{jitter_channelsep}. The total observing bandwidth is divided into 32 channels and for each frequency channel we compute a mean Spearman correlation coefficient by bootstrapping the corresponding 8~s sub-integrations. It is clear that with increasing channel separation, the ToAs begin to decorrelate, implying that the bandwidth of the process that causes jitter is comparable to the bandwidth of our observations.

\section{Limits on DM measurements in PSR J0437--4715} \label{0437_DM}

\begin{figure*}
\centering
\includegraphics[angle=0,width=0.8\textwidth]{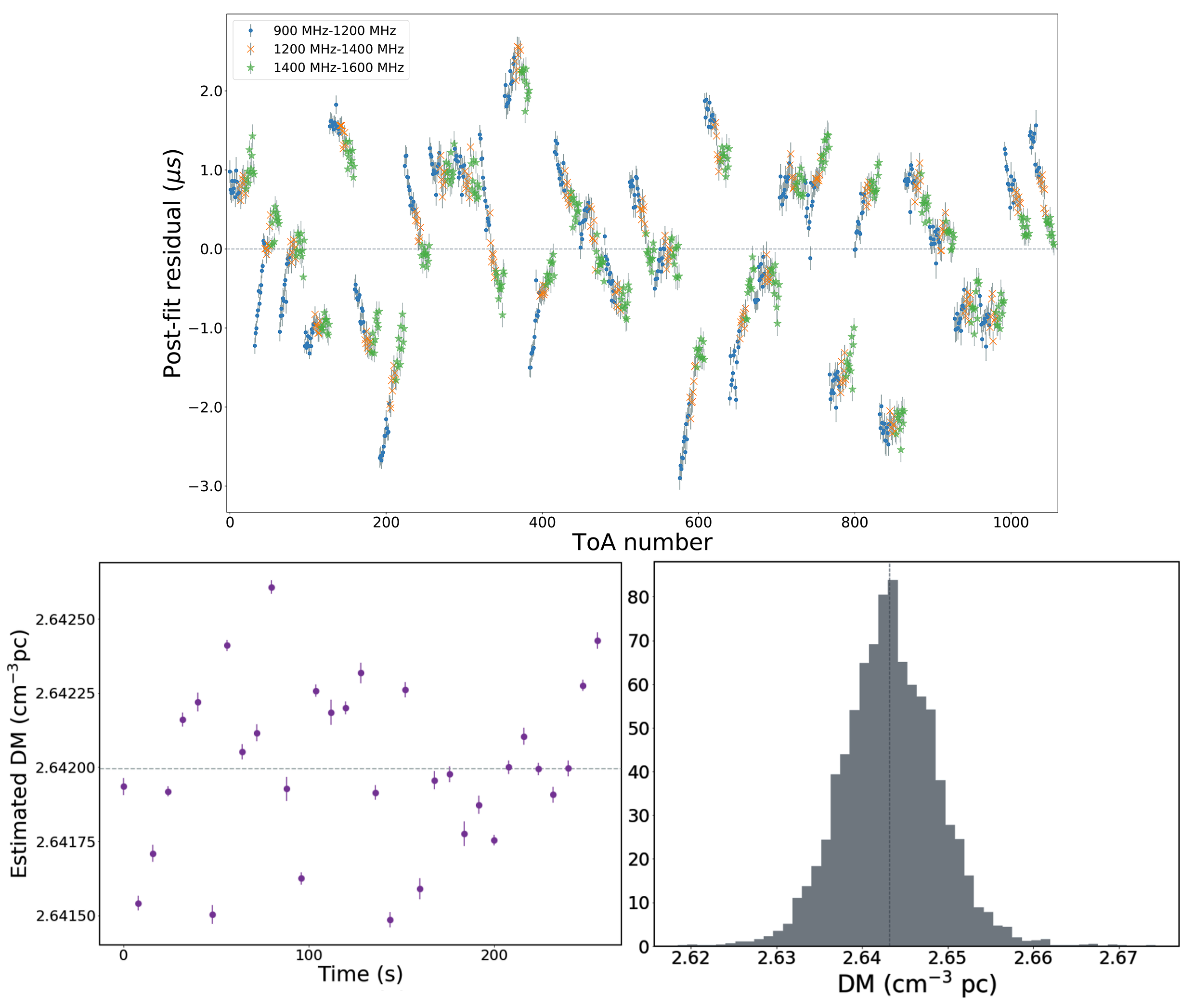}
\caption{ \label{0437_dm_plots}
\textbf{Top panel:} Post-fit wideband timing residuals of PSR J0437--4715 estimated using 8~s integrated profiles containing 32 frequency channels over a 256~s observation. Each set of ToAs are colored based on the frequency channel as indicated in the plot. The timing residuals are plotted serially (as ToA numbers) to showcase the varying frequency dependence for each ToA set. \textbf{Left panel:} Estimated values of DM for every 8~s subintegration from fits to the timing residuals shown in the top panel. The horizontal dashed line represents the median estimated DM value of 2.6419 cm$^{-3}$ pc. \textbf{Right panel:} Distributions of measured DM values from $\sim$ 47000 single pulses for PSR J0437--4715.}
\end{figure*}

The emergence of decorrelation of jitter with observing frequency in PSR J0437--4715, implies that at the lower and higher frequency bands, the emission statistics are increasingly independent of each other. In the top panel of Figure \ref{0437_dm_plots}, we show the post-fit timing residuals (generated using a wideband template) from each frequency channel plotted serially in time across a 256~second observation. Each cluster of ToAs is 8~s long with the color representative of the frequency band of observation. A striking feature is the varying frequency dependence of the arrival times on a timescale of $\sim$ 8 seconds which can naively be interpreted as a change in the DM. It must be noted that if a wideband template is not used to compute the ToAs, the dominant frequency-dependent term in the timing residuals is caused by profile evolution. An implication of this time-varying frequency dependence is that the measured values of DM appear to vary on such short timescales. Measuring the DM independently from each 8~s cluster of ToAs, we obtain a median value of 2.6419 cm$^{-3}$ pc with a standard deviation of 2.7$\times$10$^{-4}$ cm$^{-3}$ pc as shown in the left panel of Figure \ref{0437_dm_plots}. We also investigate this effect by analysing the single pulses from this pulsar. For each pulse with full frequency resolution, we compute the arrival times using a wideband template and estimate the DM. The right panel of Figure \ref{0437_dm_plots} shows the distribution of estimated DMs with a median value of 2.643 cm$^{-3}$ pc and a standard deviation of 8.5$\times$10$^{-3}$ cm$^{-3}$ pc, which is consistent with what is expected from extrapolating the 8-s subintegrations.

\begin{figure*}
\centering
\includegraphics[angle=0,width=0.8\textwidth]{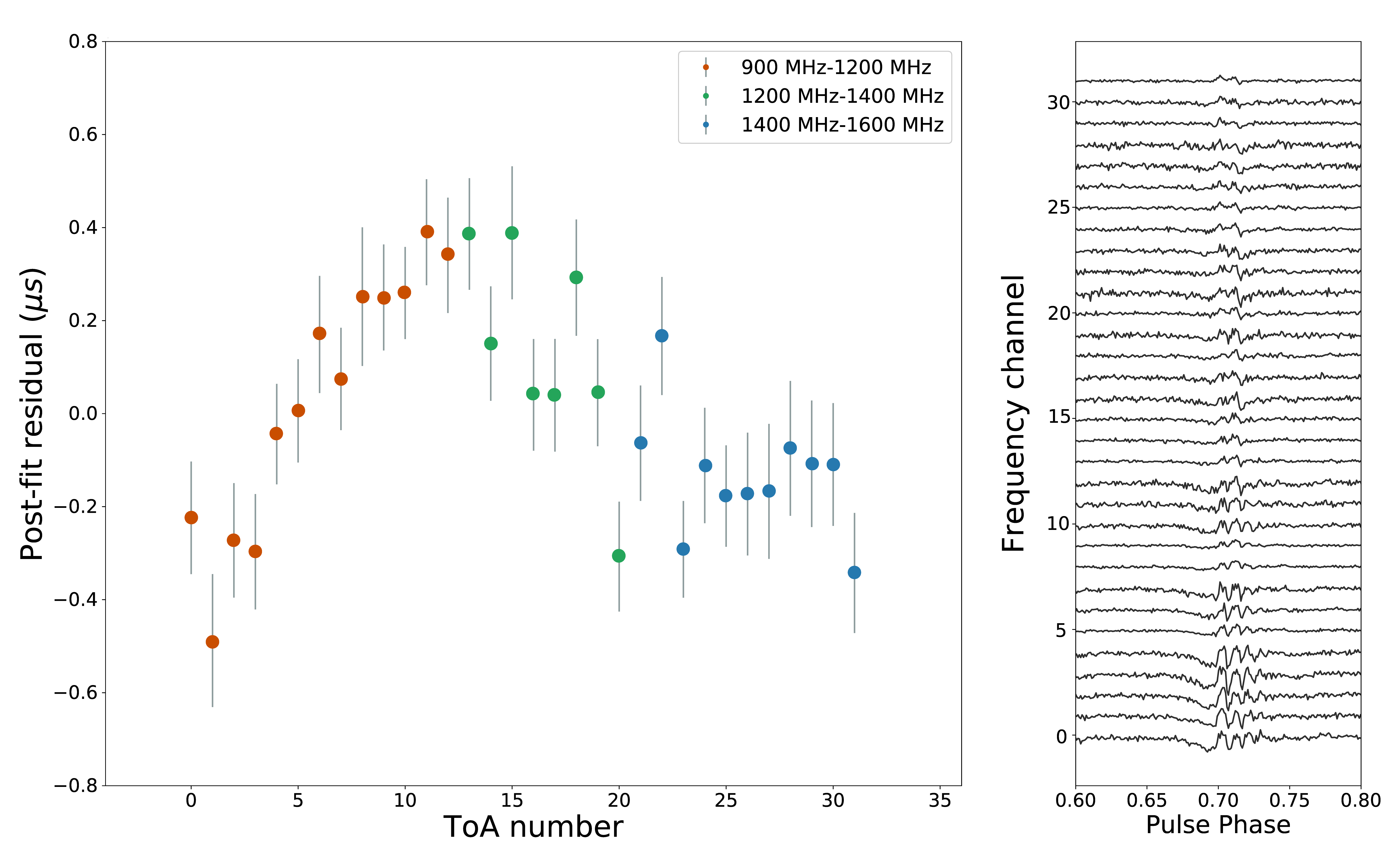}
\caption{ \label{0437_dm_profresid}
\textbf{Left panel:} Sub-banded timing residuals of a single 8~s integrated profile of PSR J0437-4715 plotted serially in time, after fitting for DM. \textbf{Right panel:} Pulse profile residuals as a function of observing frequency computed by subtracting an average pulse frequency evolution model from the same 8~s integrated profile.}
\end{figure*}

The left panel of Figure \ref{0437_dm_profresid} shows a single 8~s integration of wideband timing residuals selected from Figure \ref{0437_dm_plots} but after fitting for DM, and the right panel shows the corresponding profile residuals. The clear presence of structures in the post-fit timing residuals suggests that there may be other (possibly intrinsic) processes that produce a spectral dependence on the ToAs. It must be noted that the fit for DM also absorbs the $1/\nu^2$ contributions from such processes, thereby introducing a large scatter in the DM estimates. Analysing consecutive single pulses, we find that the characteristics of the spectral structures changes from pulse to pulse, potentially causing the varying frequency dependence as shown in Figure \ref{0437_dm_plots}. This is likely to be ``spectral jitter", a phenomena where the amount of jitter varies stochastically across the observing band. 

\begin{figure*}
\centering 
\includegraphics[angle=0,width=1\textwidth]{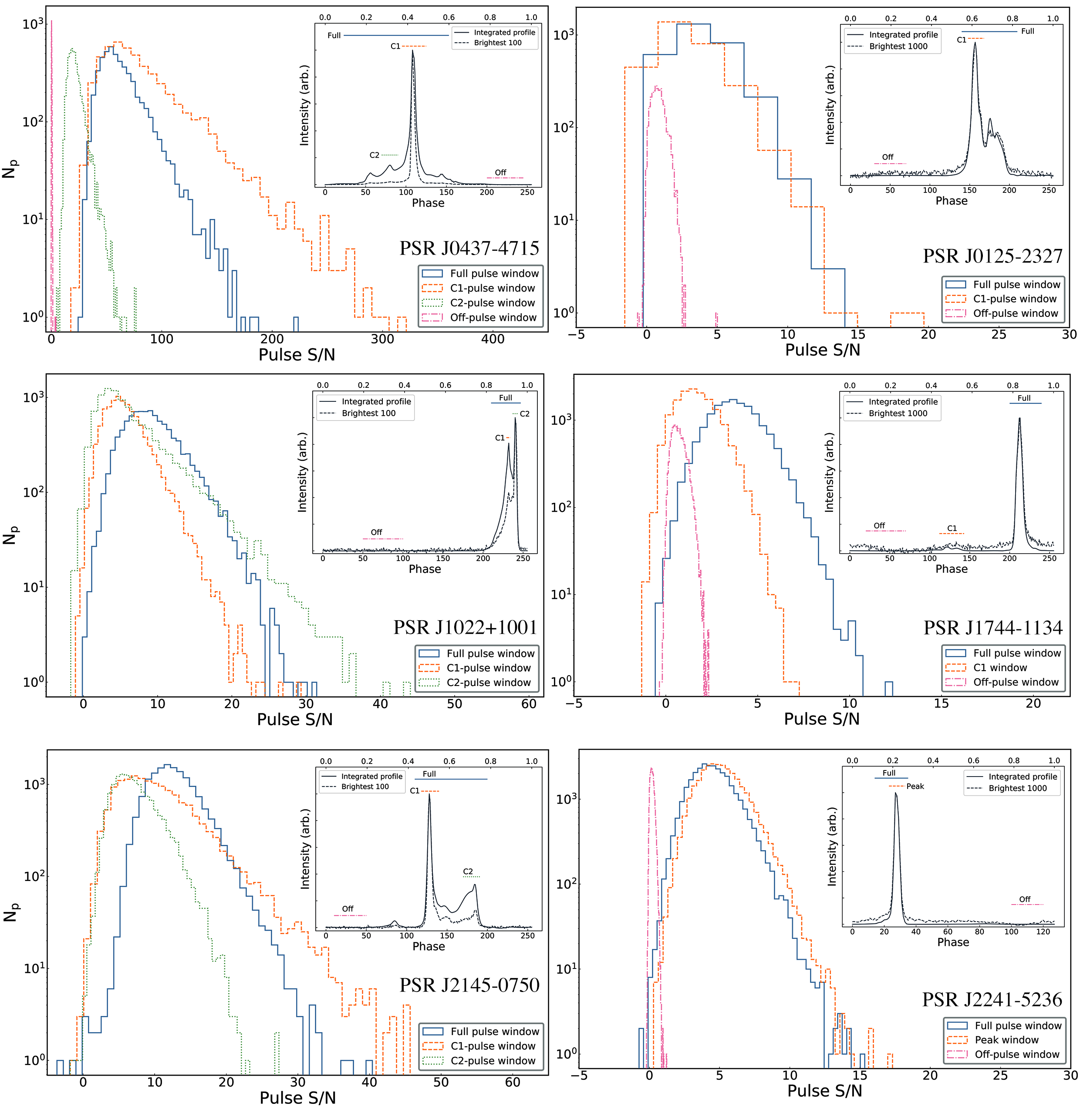}
\caption{\label{snrdist_6msps} Single pulse S/N histograms for six MSPs. The various histograms shown for each pulsar correspond to selected windows across the pulse profile. The windows used for each pulsar are shown in the respective sub-plot containing the integrated profile (solid line) and the mean profile which is derived from averaging the brightest 100 or 1000 single pulses (dashed line). The bottom axis for these sub-plots represent the number of phase bins used while the top axis shows the corresponding phase in turns. The S/N across a selected window is computed using Equation \ref{s/n_hist}.}
\end{figure*}

To understand why such a spectral structure arises, we analysed phase-resolved modulation index of the pulsar using its single pulses. We find that the modulation index of the main component (C1; see Figure \ref{snrdist_6msps}) is much higher than the other (wings) parts of the profile.
The shape  of the 8-second integrations is thus dependent on the instantaneous modulation of each profile components, which causes the profile shape to significantly deviate from the average profile.

An alternative, more speculative perspective is that the observed apparent DM variations could arise due to changes in the plasma density inside or nearby the pulsar magnetosphere. We note that the magnitude of the observed DM variations, 8.5$\times$10$^{-3}$ cm$^{-3}$ pc, would in principle allow for such a possibility. Previous work, for  instance, by \cite{Wu_pulsarDM_1994} or \cite{Luo_plasmaprocesses_1998}, have proposed that DM variations arising from non-linearities in the pulsar magnetosphere may lead to a strong dependence between radio luminosity and DM fluctuations. Since we do not find any correlation between single-pulse DM estimates and S/N and that the observed arrival time variations do not show $\nu^{-2}$ scaling, the argument for a dispersive origin for the frequency dependent variations is not strongly supported. 


In summary, it is evident that in PSR J0437--4715, spectral jitter places a fundamental limit on the precision of DM estimates on short timescales and it is likely that this phenomena can be observed in other bright nearby pulsars. 

A similar effect has been observed in PSR~J1713$+$0747. In the highest S/N observation obtained with the Arecibo telescope in a band spanning 1.1-1.7 GHz (\citealt{lam_16_intraday}), sub-banded arrival times showed temporally uncorrelated, frequency-dependent linear variations in arrival times  in both $10$-s and $80$-s sub-integrations.  In $80$ sub-integrations, the arrival time variations had a typical value of $\sim 0.3~\mu$s\,MHz$^{-1}$, which extrapolates to  $0.9\,\mu$s\,MHz$^{-1}$ for $8$\,s subintegrations.  In J0437$-$4715 $8$\,s observations we observe a typical drift\footnote{The drifts in the MeerKAT  observations of PSR~J0437$-$4715 often depart markedly from a linear trends.} of $\approx~1\,\mu$s GHz$^{-1}$.  The similarity in the values can be explained by the pulsars having comparable levels of jitter noise and pulse profile evolution.

\section{Single-pulse phenomenology} \label{singlepulse_sec}

The phenomenon of jitter can be better understood by examining shape variations of single pulses. Single pulses from MSPs have been studied in relatively few cases compared to more slower, normal pulsars, owing to their low S/N per pulse. Out of the 29 pulsars listed in Table \ref{jitter_table}, we present a statistical analysis of single pulses for eight of them with detected jitter measurements.  
Some pulsars with inferred jitter measurements (from Section \ref{jitter_measurements_sec}) did not have associated search mode observations to enable single pulse analysis and others with upper limits on jitter had very low S/N single pulses.

For each pulsar, a set of statistical properties, derived from frequency-averaged pulse profiles were determined to study and compare their emission properties. These are described below:

\textbf{Characterising variations in single pulse morphology:} 
Examining the brightest pulses and their phase-resolved modulation can provide insights into the state of plasma emission. We examine a number of statistical properties to characterise single pulse amplitude and shape variability.

    We compute the phase-resolved modulation index as,
    \begin{equation} \label{mi_eq}
    m_{I}(\phi)=\frac{\sqrt{\sigma_{I}^{2}(\phi)-\sigma_{ \mathrm{off}}^{2}}}{I(\phi)},
    \end{equation}
    \noindent where $\sigma_{I}(\phi)$ and $I(\phi)$ are the rms and the mean intensity computed at phase $\phi$ and $\sigma_{ \mathrm{off}}$ is the rms intensity computed from an off-pulse window. 
    
    We also investigate temporal correlations in intensities of single pulse emission,  in particular to check whether emission shows the pulse-pulse correlation observed in many young pulsars \cite[e.g.,][]{Backer_1970_Modechanging}.  We implemented this through analysis of the time series of maximum intensity of the individual pulsars. In particular we calculated the auto-correlation function (ACF) of the time series.

\textbf{Pulse energy distributions:} 
Analysis of single pulse energy distributions provide a measurement of energy contained in a pulse or sub-component(s) of a pulse and the type of distribution allows us to probe the pulse emission mechanism. We measure the integrated S/N by defining windows around the main component and various sub-components of the main pulse and interpulse depending on the profile morphology. In cases where multiple sub-components were present, the size of the windows were chosen to be similar (wherever reasonable) to enable a more direct statistical comparison. The instantaneous S/N over a selected window is computed as, 
    \begin{equation} \label{s/n_hist}
    S / N=\frac{\sum_{i=1}^{N_{\text {window }}}\left(A_{i}-B\right)}{\sqrt{N_{\text {window }}} \sigma_{\text {off }}},
    \end{equation}
    \noindent where $A_{i}$ is the pulse flux density at the $i$-th bin, $B$ is the mean off-pulse flux density, $N_{\rm window}$ is the number of phase bins in the selected component window and $\sigma_{\rm off}$ is the off-pulse rms flux density. RFI excision on single pulses is implemented using standard \textsc{psrchive} tools and \textsc{coastguard}, whenever necessary. Residual RFI was manually inspected and removed from the analysis. A least-squares minimisation of the data, fitted by a model for the pulse energy distribution was performed using log-normal and Gaussian distributions. The log-normal model was defined as,
    
    \begin{equation}
        P(S)=\frac{1}{S \sigma_{\ell} \sqrt{2 \pi}} \exp \left[\frac{-\left(\log _{10}(S)-\mu_{\ell}\right)^2}{2 \sigma_{\ell}^{2}}\right],
    \end{equation}
    
    \noindent where $S$ is the S/N , and $\mu_{\ell}$ and $\sigma_{\ell}$ parameterise the distribution. For the Gaussian model, we fit for the mean ($\mu_{g}$) and standard deviation ($\sigma_{g}$) of the pulse energy distribution. A $\chi^{2}$ statistic was used for quantifying the goodness of fit of the preferred distribution. 
    
 \textbf{Timing properties of single pulses:} 
By computing frequency-averaged single pulse ToAs and investigating correlations between single pulse properties and corresponding ToAs, we study the timing properties of single pulses. We also measure jitter as a function of the number of pulses integrated ($N_{\rm p}$) and show that the arrival time uncertainties obtained from averaged profiles due to template fitting are consistent with the measured pulse-to-pulse variations and that jitter typically scales as $1/\sqrt{N_{\rm p}}$.


\begin{figure*}
\centering
\includegraphics[angle=0,width=1\textwidth]{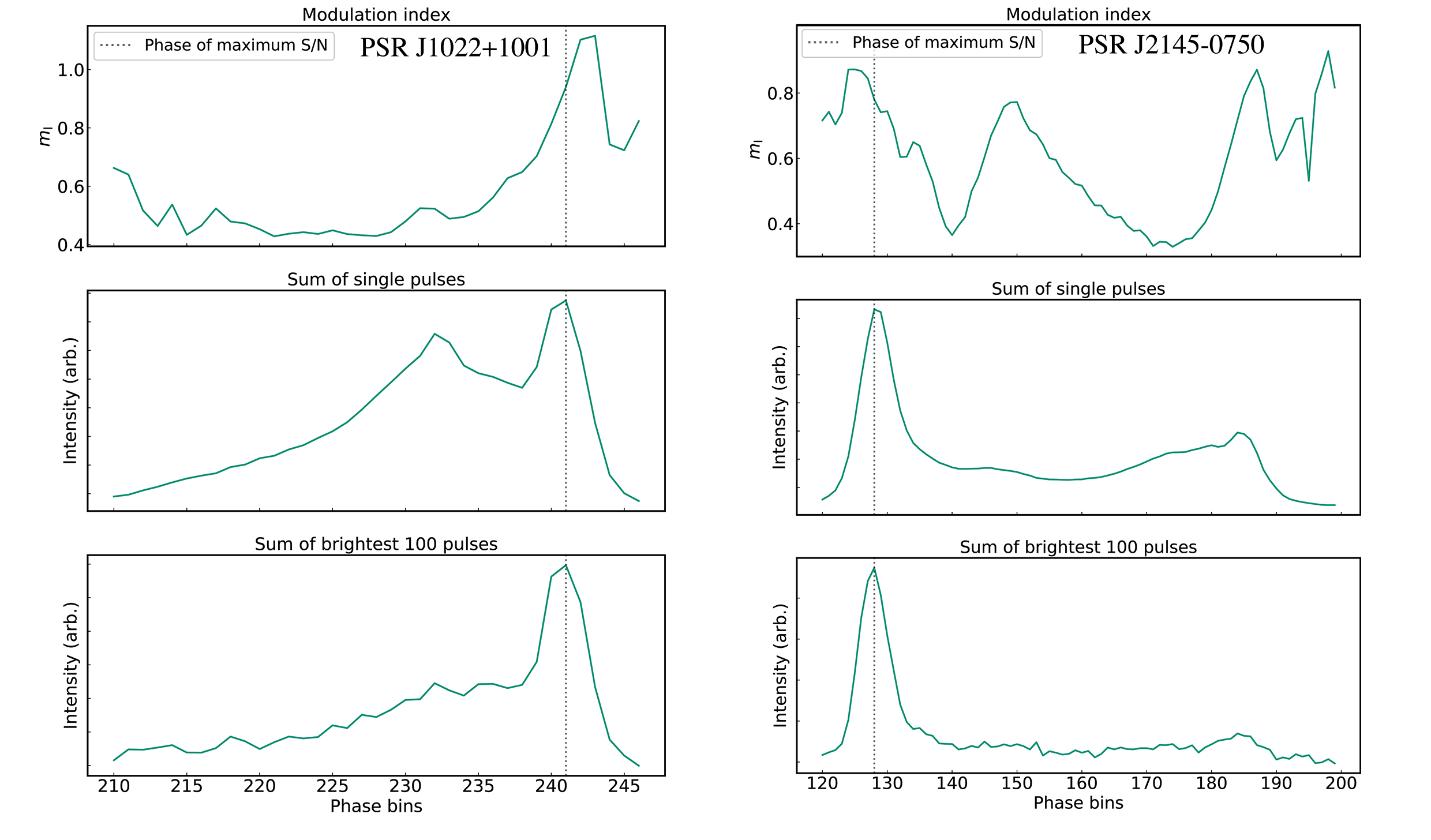}
\caption{\label{mi_plots} Phase-resolved modulation index for PSRs J1022+1001 and J2145--0750 shown across the on-pulse region in the upper panel. The middle panel shows the integrated profile from 15000 single pulses and the lower panel shows the mean profile from averaging the brightest 100 pulses. The vertical grey dashed line represents the phase of maximum S/N for the profile shown in the lower panel.}
\end{figure*}

\subsection{PSR J0437--4715}

The single pulses of PSR J0437--4715 have been studied extensively and owing to its high flux density, pulse shape variations cause excess timing uncertainty, at least four times greater than that predicted from radiometer noise alone (\citealt{stefan_sp}). We analysed $\sim$ 47000 pulses and detected every single pulse (the lowest S/N was $\sim$ 20), indicating that the pulse emission in this pulsar likely does not exhibit nulling. Its profile has multiple components that span the majority of the pulse phase. The S/N distribution was analysed by selecting different windows spanning the pulse profile as shown in Figure \ref{snrdist_6msps}. The profile formed by averaging the brightest 100 pulses is shown in the subplot along with the integrated profile. The S/N histograms for the C1, C2 and the full profile windows show log-normal distributions. It can also be seen that the brightest pulses coincide in phase with the leading edge of the integrated profile and that the brightest pulses are narrower than the rest and do not exhibit strong emission in the wings of the profile. The phase-resolved modulation index is highest towards the leading edge of the main component, consistent with previous results (\citealt{stefan_sp})\footnote{The off-pulse variance was not subtracted to unbias the measured modulation index in \cite{stefan_sp}}. We computed the ACF from the peak flux intensities of single pulses and find no evidence of temporal correlations amongst the pulses. The level of jitter noise estimated by integrating increasing number of pulses from 10 to $\sim$10000 shows that jitter scales proportionally with 1$/\sqrt{\rm N_{\rm p}}$ and is consistent within uncertainties over multiple epochs. Our statistical analysis of single pulses from PSR J0437--4715 show that they are consistent with previously published analyses (\citealt{jenet_2001}, \citealt{stefan_sp}, and \citealt{Shannon_jitter_2014}).

\subsection{PSR J0125--2327}

PSR J0125--2327 is a binary pulsar with a spin-period of $\sim$ 3.6~ms, discovered in the Green Bank North Celestial Cap (GBNCC) survey (\citealt{GBNCC_2020}). With MeerKAT, the observed median S/N per pulse was $\sim$4 with an estimated $\sigma_{\rm J}$ to be 48$\pm$13 ns in an hour. Statistical properties of single pulses from PSR J0125--2327 have not been previously reported. A total of $\sim$ 15000 pulses were analysed and the maximum observed S/N was $\sim$ 15 per pulse. In the observed band, the pulse profile has multiple components with a main strong component towards the leading edge of the profile and multiple weaker components towards the trailing edge. The distribution of S/N computed from single pulses follows a log-normal distribution as shown in Figure \ref{snrdist_6msps} for two selected windows over the full on-pulse region and the peak component (C1). Unlike PSR J0437$-$4715, bright pulses from this pulsar span the entire on-pulse phase region. 





\subsection{PSR J1022+1001}

PSR J1022+1001 is a relatively bright MSP with a rotation period of $\sim$ 16~ms. Owing to its rotational stability and low mean arrival time errors, it is a part of the Parkes Pulsar Timing Array program (PPTA) (\citealt{ppta}). Previous measurements of pulse shape variations have shown that there is excess scatter in ToAs, larger than that expected from only radiometer noise (\citealt{Kuo_1022_2015}). The pulse profile at 20 cm wavelengths has a double peaked structure and each component has a different spectral index, resulting in strong evolution of the pulse profile across the observing band. We analysed $\sim$ 20000 pulses and found the maximum S/N to be $\sim$ 30. In the left panel of Figure \ref{mi_plots}, we show the phase-resolved modulation index computed across the on-pulse region. The modulation index is a measure of intensity variations from pulse to pulse. In this pulsar, the modulation index grows increasingly stronger across the pulse profile and becomes strongest towards the trailing edge of the second peak component suggesting high levels of amplitude modulation in the pulses originating from these phase ranges. The modulation index of the first component is $\sim$ two times smaller than the second component. The brightest pulses are dominated by emission from the second peak component and coincide in phase with the trailing edge of the second component as shown by the vertical dotted lines. From the ACF of pulse peak flux intensities, we find no evidence of temporal correlations amongst pulses. The single pulse S/N distributions of the pulse profile and its peak components are shown in Figure \ref{snrdist_6msps} and follow a log-normal distribution. While the integrated pulse profile formed from 100 brightest pulses shows evidence of both the components, the second component (C2) is much more prominent than the first (C1) which is also supported by the S/N distributions which show that the brightest pulses originate from C2. We estimate $\sigma_{\rm J}$ to be 130$\pm$20 ns in an hour for PSR J1022+1001 and find that our measurements of jitter scale as $1/\sqrt{N_{\rm p}}$ which are consistent over multiple epochs.



\subsection{PSR J1603--7202}
PSR J1603--7202 has a spin period of $\sim$ 15~ms and at 20 cm wavelengths has two sub-components connected by a more dominant bridge of emission. We analysed $\sim$ 20000 pulses and found the maximum S/N to be $\sim$ 17. Unlike in PSR J1022+1001, the modulation index of this pulsar is strongest towards the leading edge of first component and is much weaker towards the second component. The S/N distribution of both the components follow a log-normal distribution. We estimate $\sigma_{ J}$ to be 180$\pm$40 ns in an hour and find that it scales as $1/\sqrt{N_{\rm p}}$ similar to other pulsars. 




\subsection{PSR J1744--1134}
PSR J1744--1134 has a spin period of $\sim$ 4~ms and has a relatively narrow main pulse profile with an interpulse at 20 cm wavelengths. We analysed $\sim$ 35000 single pulses and found the maximum S/N to be $\sim$ 12. Unlike pulsars discussed so far, the S/N distributions for both the peak pulse component and the interpulse tends towards a Gaussian distribution consistent with low amplitude modulation as shown in Figure \ref{snrdist_6msps}. In contrast to PSR J1022+1001, where the modulation index peaks towards the trailing edge of the second component, in PSR J1744--1134, it gets weaker towards the peak component implying low levels of amplitude modulation. We do not detect single pulse emission from the interpulse (C1). We estimate $\sigma_{\rm J}$ to be 30$\pm$6 ns in an hour and similar to other pulsars it scales as $1/\sqrt{N_{\rm p}}$.   

    

\subsection{PSR J1909--3744: pulse nulling?}

\begin{figure*}
\centering
\includegraphics[angle=0,width=1\textwidth]{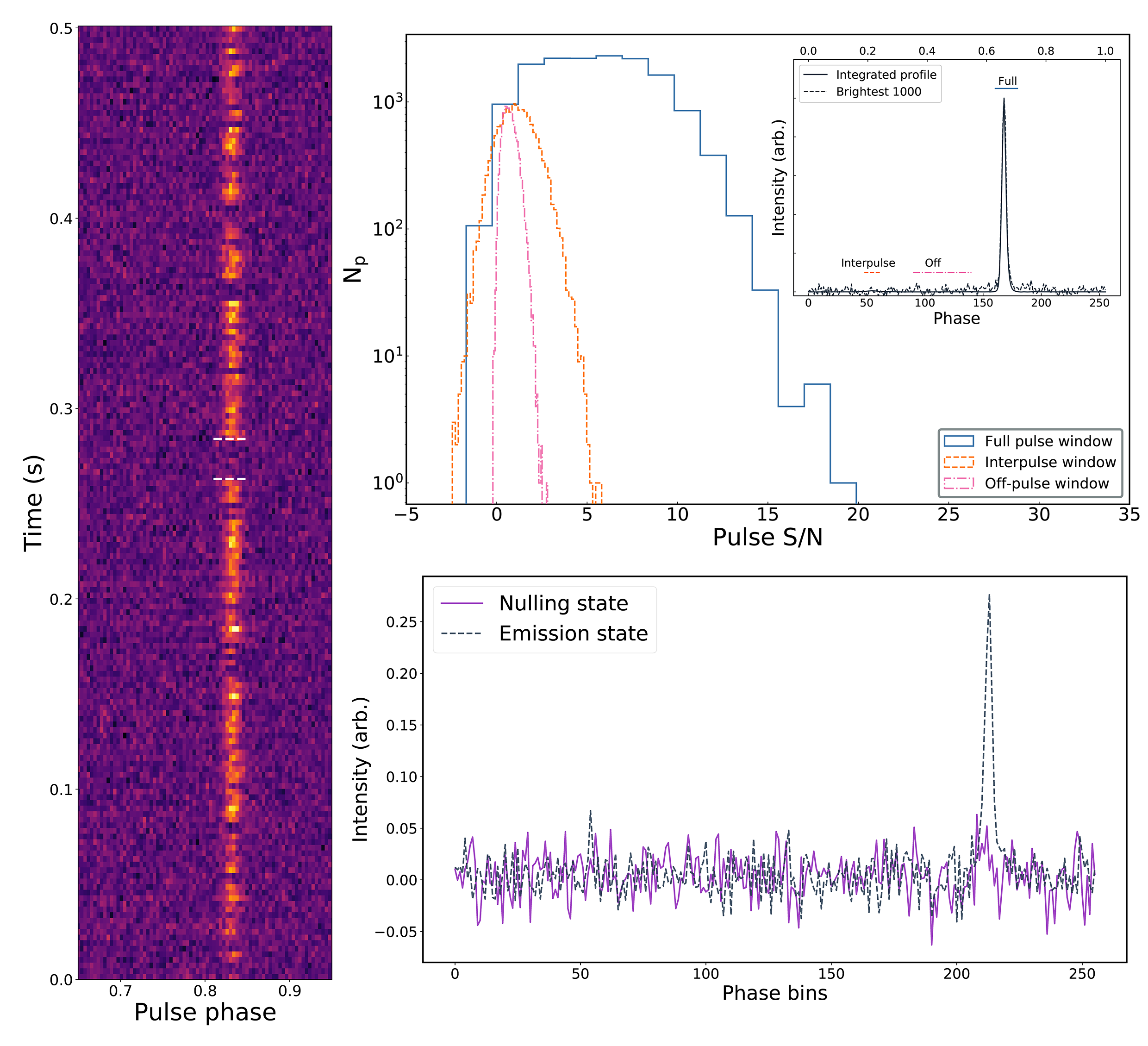}
\caption{\label{1909_plots} \textbf{Left panel:} Consecutive single pulses of PSR J1909--3744 shown across the on-pulse region indicating potential pulse nulling behaviour as depicted by the two horizontal dashed lines. \textbf{Right top panel:} Single pulse S/N histograms for PSR J1909$-$3744 similar to Figure \ref{snrdist_6msps}. We note that the window sizes for the interpulse and the off-pulse are different which leads to a broader S/N distribution for the interpulse relative to that of the off-pulse. \textbf{Right bottom panel:} Profile formed from integrating the amplitudes of eight consecutive pulses over the region highlighted in the left panel is shown in purple relative to an integrated profile formed from adding eight pulses in the emission region (black dashed line) for reference.}
\end{figure*}

\begin{figure*}
\centering
\includegraphics[angle=0,width=1\textwidth]{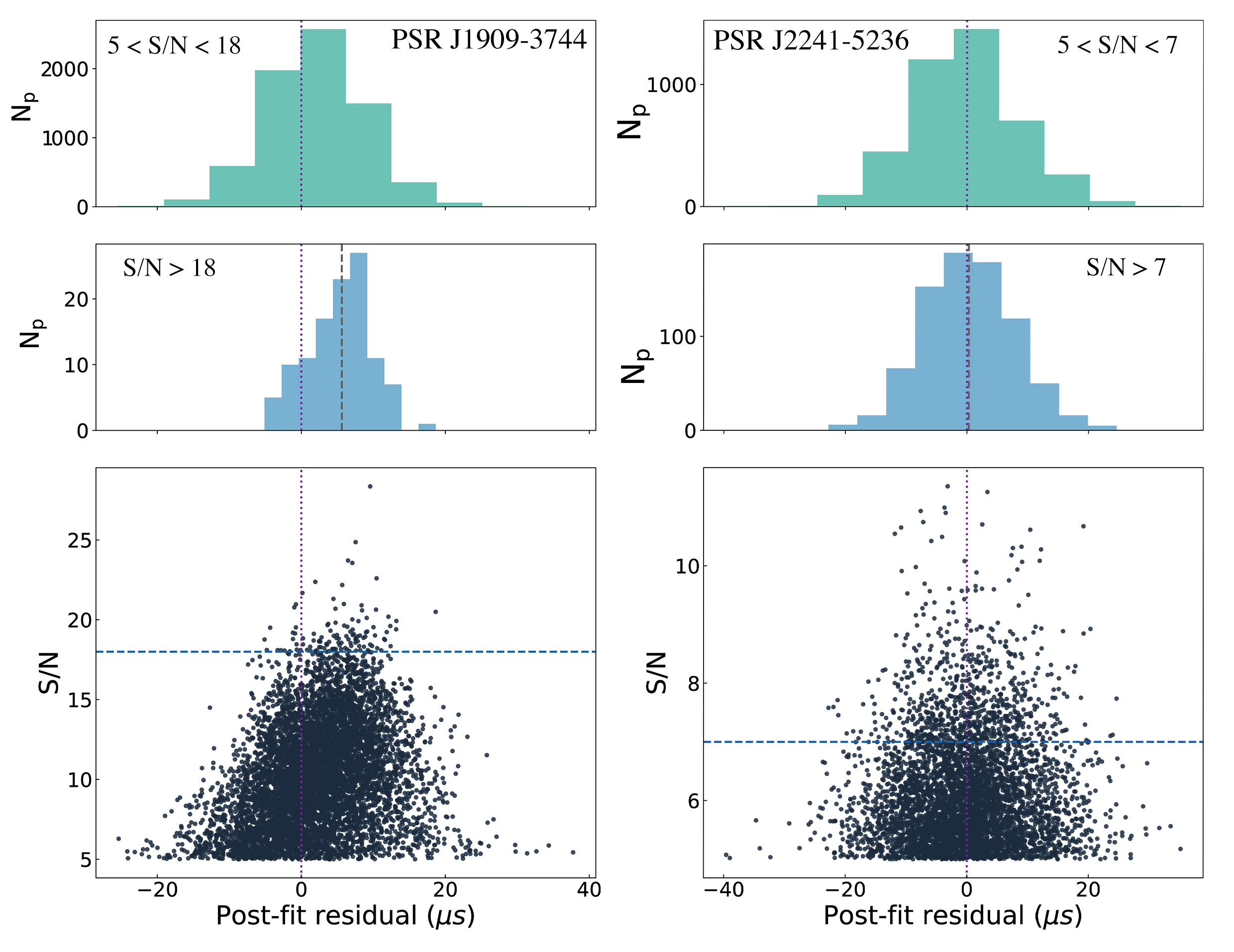}
\caption{\label{toasnr_plots} ToAs computed from a randomly chosen subset of $\sim$ 10000 single pulses plotted against their corresponding S/N for PSRs J1909--3744 (left) and J2241-5236 (right). The blue dashed horizontal line represents the S/N threshold for selecting the brightest pulses which are indicated in the top and middle panels. The vertical dashed line in the middle panel represents the median of the histogram, while the vertical dotted line across all three panels is at zero $\mu s$.}
\end{figure*}

PSR J1909--3744 is one of the most precisely timed pulsars owing to its extremely stable rotational behaviour, narrow pulse profile and high flux density. At 20 cm wavelengths, the pulse profile consists of a narrow main component with an interpulse. We analysed $\sim$ 35000 pulses and found the maximum S/N to be $\sim$ 20 but do not detect single pulses from the much fainter interpulse. In the left panel of Figure \ref{1909_plots}, we show $\sim$ 200 consecutive pulses in which the pulse emission appears to null occasionally as highlighted by two horizontal dashed lines at $\sim$ 0.28 seconds. This is the first reported detection of a nulling phenomenon in an MSP. Integrating the flux densities over the nulling region reveals no detection of emission as shown in the bottom right panel. However, it must be noted that analysing longer observations using high time resolution search-mode data might help identify any weak emission modes. The S/N distribution of single pulses of the main component appears to follow a Gaussian distribution similar to PSR J1744$-$1134 although with a broad spread about the mean value (i.e. a platykurtic distribution) as shown in the top right panel. It is also interesting to note that the mean profile formed from integrating the brightest 100 pulses has relatively more flux density towards its leading edge as compared to the integrated profile resulting in the brightest pulses (S/N $>$ 18) having later times of arrival than pulses with average S/N (5 $<$ S/N $<$ 18) as shown in Figure \ref{toasnr_plots}. We estimate the value of $\sigma_{\rm J}$ to be 9$\pm$3 ns in an hour and find that when integrating from 10 to 10000 pulses, jitter scales as $1/\sqrt{N_{\rm p}}$.

    


\subsection{PSR J2145--0750}
PSR J2145--0750 has a spin period of $\sim$ 15~ms and is a relatively bright pulsar at 20 cm wavelengths. Its pulse profile is dominated by two main components connected by a bridge of emission and a precursor. We detect single pulses from both the main components and none from the precursor. We analysed $\sim$ 16000 pulses and found the maximum S/N to be $\sim$ 50. The modulation index shows complex structures, with high levels of modulation towards the leading edge of the first component and the trailing edge of the second component with the bridge also exhibiting a high modulation index as shown in the right panel of Figure \ref{mi_plots}. The S/N histograms of the two components (C1 and C2) and the full on-pulse window show log-normal distributions as shown in Figure \ref{snrdist_6msps}. The wide variability in phase and amplitude exhibited by the single pulses results in large levels of jitter noise. We estimate a $\sigma_{\rm J}$ value of 200$\pm$20 ns in an hour and find that jitter scales as $1/\sqrt{N_{\rm p}}$. 




\subsection{PSR J2241--5236}

PSR J2241--5236 has a spin period of $\sim$ 2.2~ms and is in a 3.5 hour orbit with a low mass companion (\citealt{keith_2011_mspdiscoveries}). The higher flux density, low dispersion measure and rotational stability makes it an excellent candidate for high precision pulsar timing experiments. Consecutive single pulses from PSR J2241--5236 exhibit very high levels of stability in phase. Its pulse profile at the observed band has a narrow main component with an interpulse, similar to PSR J1909--3744. We analysed $\sim$ 24000 pulses and found the maximum S/N to be $\sim$ 15. We did not detect any single pulses from the interpulse. The S/N histograms of the single pulses follow an approximately Gaussian distribution as shown in Figure \ref{snrdist_6msps}. Its modulation index is weakest towards the peak of the pulse profile, similar to PSR J1909--3744 indicative of low levels of amplitude modulation. Accordingly, this pulsar exhibits the lowest levels of jitter hitherto reported of just 3.8$\pm$0.8 ns in an hour. The brightest pulses (S/N $>$ 7) largely resemble the integrated profile implying that their ToAs are expected to have similar arrival times as pulses with average S/N (5 $<$ S/N $<$ 7) as shown in Figure \ref{toasnr_plots}. 

The S/N of single pulses were not sufficient to estimate the scaling relation of jitter. The observation that was used to measure jitter from the folded archives (in Section \ref{jitter_measurements_sec}) had an estimated median single pulse S/N of $\sim$ 10. However, there was no associated single pulse data for that particular observation. In comparison, the median single pulse S/N that are presented here are only $\sim$ 4. Analysis of single pulses during scintillation maxima would prove very interesting for precision timing analysis. 



\section{Discussions and Conclusions} \label{conclusion_sec}

\subsection{High-precision pulsar timing with MeerKAT}
We have presented the first short-term high-precision pulsar timing results using the MeerKAT radio telescope including a study of single pulse phenomenology for a selection of MSPs. We found that in the highest S/N observations, stochastic pulse shape and amplitude variations cause excess scatter in the ToAs, which are higher than that expected from radiometer noise alone. Out of the 29 MSPs in our sample, we reported new jitter measurements for 15 pulsars, out of which six pulsars had constraint measurements while upper limits were reported for the remaining. In the remaining 14 pulsars, we found that our measurements are either consistent with previously reported values or have tighter upper limits. PSR J2241--5236 has the lowest levels of jitter reported in any pulsar of just $\sim$ 4~ns in an hour followed by PSR J1909--3744 with a jitter level of $\sim$ 9~ns in the same time. The upper limit on PSRs J1017--7156, J1946--5403 and J2129--5721 is $<$ 11~ns in an hour,  which promise to be excellent pulsars for high-precision timing experiments. New detections of jitter in MSPs clearly confirm  that jitter is a generic property in all pulsars and our ability to detect jitter in many more pulsars will grow with increasing sensitivity of radio telescopes. It also shows that the levels of jitter vary markedly across the population, even between pulsars with similar spin periods and pulse profiles.  

We find evidence for the frequency dependence of jitter in PSR J0437--4715 and showed that jitter decreases only moderately with increasing observing frequency. We also found that jitter decorrelates over the observing bandwidth of MeerKAT implying that the pulse emission statistics will become increasingly independent to each other with wider bandwidths.
This is consistent with analysis of the pulsar obtained with the  64-m Parkes radio telescope, which showed no correlation in jitter in observations obtained at bands centred at $730$~MHz and $3100$~MHz \cite[][]{Shannon_jitter_2014}.
The decorrelated nature of emission also results in individual pulses having spectral structures that vary from pulse-to-pulse and owing to the very high S/N of each pulse across the observing band, we detect this effect in the pulsar timing residuals as time-varying frequency dependence. Measurements of DMs on short-timescales will thus have a large scatter, placing a fundamental limit on the precision with which pulsar DMs can be measured. In addition to the limitations posed to the measurement of the gravitational wave background signal, this effect can also lead to limiting the precision of orbital parameters in highly relativistic systems, especially due to covariances between the DM and timing model parameters when combining observation from multiple orbits. 

We speculate that with more sensitive radio telescopes with larger bandwidths, pulse jitter will dominate the error budget. 


\subsection{Pulse shape variability in MSPs}
We reported the first single pulse study of eight MSPs using MeerKAT. Single pulse statistics for PSRs J0125-2327 and J2241--5236 have not been previously studied. We also reported the first observation of pulse nulling in an MSP, seen in PSR J1909--3744. Most pulsars in our sample with detected single pulses showed log-normal distributions consistent with previous results with MSPs (\citealt{Shannon_jitter_2014}) and slow spinning pulsars (\citealt{Burke_Spolaor_Singlepulses}). However, pulsars with the lowest levels of jitter noise, PSRs J2241--5236, J1909--3744 and J1744--1134 showed approximately Gaussian distributions.


 Studying energy distributions enables us to distinguish between different models of plasma behaviour based upon theoretical predictions. For example, self-organised criticality (\citealt{bak_soc,bak_soc1}) describes self-consistent systems that interact without any preferred distance or timescales. It predicts a power-law distribution of pulse intensities. However, models based on stochastic growth theory (\citealt{robinson_sgt,robinson_sgt2}) describe interactions in an independent homogeneous medium with preferred distances and timescales. This model predicts a log-normal distribution of pulse intensities. \cite{Cairns_Johnston_Das_2004} found Gaussian energy distributions at the edges of the pulse profile in slow spinning pulsars. They attribute this to either irregularities in the density of the plasma or a superposition of multiple weak emission components. Following from this, it can perhaps be speculated that for the three pulsars in our sample showing Gaussian energy distributions, our line of sight traverses the edge of the emission region. It is also interesting to note that these three pulsars have single component Gaussian profiles with an interpulse.  
 
 In pulsars that show high levels of jitter noise, we found that the phase-resolved modulation index increased towards the main emission component, such as in PSRs J0437--4715, J1022+1001, J1603--7202 which have wide profiles with multiple components; whereas in pulsars with the lowest levels of jitter noise, the modulation index showed the opposite trend and approached a minima towards the main emission component, such as in PSRs J1744--1134, J1909--3744 and J2241--5236 which have narrow profiles with single components and a weak interpulse. 
 
 The phase-resolved modulation index can be used to distinguish between pulsar emission models and previous theoretical works have suggested that the modulation index depends upon some function of the pulsar period and its period derivative. It is however, somewhat arbitrary to define a modulation index from the phase-resolved modulation index profile because it is highly dependent on the pulse phase. In this case, we chose a value that is representative of the modulation index near the main emission component of the integrated pulse profile. We used a Spearman correlation coefficient to determine the correlations between the modulation index and other pulsar parameters and found moderate correlations of 0.62$\pm$0.08 and 0.54$\pm$0.03 between the period, period derivative and the modulation index of the pulsar respectively. These measurements are consistent with previous reports (\citealt{Jenet_Gil_2003, jenet_gil_2004, Edwards_stappers_2003}) which may imply that such a relationship is a consequence of the `sparking gap' model first theorized by \cite{emission_ruderman}. 
 
 Future observations of more pulsars will enable us to identify correlations with the various `complexity parameters' (\citealt{Gil_Sendyk_2000, Lou_2001}) that aid in distinguishing between different emission models which will prove powerful in probing deeper into understanding pulsar emission models. We also measure the correlation of $\sigma_{\rm J}$ (measured in 1~hr) with various estimates of the pulse widths at 50\% ($W_{\rm{50}}$), 10\% ($W_{\rm{10}}$) and the effective pulse width ($W_{\rm{eff}}$). We estimate the effective pulse width following the definition in \cite{Downs_timing} and \cite{cordes_shannon_2010},

\begin{equation}
W_{\mathrm{eff}}=\frac{\Delta \phi}{\sum_{i}\left[P\left(\phi_{i+1}\right)-P\left(\phi_{i}\right)\right]^{2}},
\end{equation}

\noindent where $\Delta\phi$ is the phase resolution of the pulse profile in units of time and $P$ is the pulse profile which is normalised to a peak intensity of unity. We find a moderate correlation of 0.64$\pm$0.09 with $W_{\rm 50}$ and no significant relationship with $W_{\rm eff}$. These results provide excellent confirmation to those reported in \cite{lam_jitter_2019} and suggest that the level of pulse jitter does not significantly depend on the `sharpness' of the pulse profile.

\subsection{Contributing to the PTA network}
 Although the MeerKAT MSP data set is not yet sensitive to gravitational wave radiation due to its short timing baselines, there are a number of ways in which it can contribute to high precision pulsar timing and PTA experiments. Limitations in precision timing experiments are typically due to an incomplete understanding of systematics in both the instrument and analysis methodologies or due to lack of sensitivity. Precise estimates of DM variations are crucial for PTA experiments, which can be improved by conducting near simultaneous global observing campaigns of a few MSPs with telescopes like MeerKAT, GBT, Parkes, Effelsberg and the Five Hundred Meter Aperture Spherical Radio Telescope. Such campaigns, as was conducted previously on PSR J1713$+$0747 in 2013 \cite[][]{Dolch_global} not only place a strong constraint on DM variations but also help characterize the instrumental response and post-processing pipelines. With new upcoming receivers such as the Ultra High Frequency (UHF) receiver (580 MHz to 1015 MHz) and the S-band receiver (1750 MHz to 3500 MHz) (\citealt{mpisband}) at MeerKAT, high sensitivity can be achieved owing to steep spectral indices of pulsars and by dynamically observing pulsars at their scintillation maxima. 

The results presented here are very relevant to current and future radio telescopes in the context of precision pulsar timing experiments and in optimizing observing strategies. For pulsars that are limited by jitter noise, it would be better to sub-array MeerKAT to observe multiple pulsars simultaneously over longer durations than use the full array to observe pulsars one by one. The pulsar timing backend at MeerKAT has the potential to form up to four tied-array beams on the sky leading to a significant improvement in the efficiency of the timing program. Sub-arrays consisting of a small number of antennas ($\sim$ 4 to 8) could be deployed to scan the skies to look for pulsars in a scintillation maxima, which could enhance the sensitivity of pulsar timing observations and aiding high precision pulsar timing experiments. 
In the context of the IPTA, observations could potentially be optimised to have sensitive telescopes avoid observing pulsars that are jitter limited and instead  observe  relatively faint pulsars to improve the overall network sensitivity \cite[][]{kj_optimisation}.   
High precision pulsar timing will greatly benefit from utilizing MeerKAT in conjunction with established radio telescopes and observing programs from around the globe.

\section{Acknowledgements}
The MeerKAT telescope is operated by the South African Radio Astronomy Observatory, which is a facility of the National Research Foundation, an agency of the Department of Science and Innovation. This work made use of the gSTAR and OzSTAR national HPC facilities. gSTAR is funded by Swinburne and the Australian Government Education Investment Fund. OzSTAR is funded by Swinburne and the National Collaborative Research Infrastructure Strategy (NCRIS). This work is supported through Australian Research Council (ARC) Centre of Excellence CE170100004. A.P. acknowledges support from CSIRO Astronomy and Space Science. R.M.S. acknowledges support through ARC grant CE170100004. M.B, S.O, and R.M.S. acknowledge support through ARC grant FL150100148. R.M.S. also acknowledges funding support through Australian Research Council Future Fellowship FT190100155. FA gratefully acknowledge support from ERC Synergy Grant “BlackHoleCam” Grant Agreement Number 610058. T.T.P. is supported through a NANOGrav Physics Frontiers Center Postdoctoral Fellowship from the National Science Foundation Physics Frontiers Center Award Number 1430284. This work also made use of standard Python packages (\citealt{numpy}, \citealt{scipy}, \citealt{pandas}, \citealt{matplotlib}), Chainconsumer (\citealt{chainconsumer}) and Bokeh (\citealt{bokeh}).

\section*{Data Availability}
The data underlying this article will be shared on reasonable request to the corresponding author.




\bibliographystyle{mnras}
\bibliography{jitter}




\bsp	
\label{lastpage}
\end{document}